\documentclass[showpacs,preprintnumbers,floatfix]{revtex4}

\usepackage{graphicx}

\usepackage{amsmath}

\usepackage{amsfonts}

\def\be{\begin{equation}}
\def\ee{\end{equation}}
\def\bea{\begin{eqnarray}}
\def\eea{\end{eqnarray}}

\begin{document}


\title{Alignment Tests for low CMB multipoles}

\author{L. Raul Abramo}

\email{abramo@fma.if.usp.br}

\affiliation{Instituto de F\'{\i}sica,
Universidade de S\~ao Paulo \\
CP 66318, CEP 05315-970 S\~ao Paulo, Brazil}

\author{Armando Bernui}

\email{bernui@das.inpe.br}

\author{Ivan S. Ferreira}

\email{ivan@das.inpe.br}

\author{Thyrso Villela}

\email{thyrso@das.inpe.br}

\author{Carlos Alexandre Wuensche}

\email{alex@das.inpe.br}

\affiliation{Divis\~ao de Astrof\'{\i}sica,
Instituto Nacional de Pesquisas Espaciais\\
Av. dos Astronautas, 1.758, CEP 12227-010, S\~ao Jos\'e dos
Campos, Brazil}

\date{\today}

\begin{abstract}

We investigate the large scale anomalies in the angular distribution of 
the cosmic microwave background radiation as measured by WMAP using 
several tests. 
These tests, based on the multipole vector expansion, measure correlations 
between the phases of the multipoles as expressed by the directions of the
multipole vectors and their associated normal planes. 
We have computed the probability distribution functions for 46 such tests,
for the multipoles $\ell=2-5$. 
We confirm earlier findings that point to a high level of alignment between 
$\ell=2$ (quadrupole) and $\ell=3$ (octopole), but with our tests we do not 
find significant planarity in the octopole. In addition, we have found other 
possible anomalies in the alignment between the octopole and the $\ell=4$ 
(hexadecupole) components, as well as in the 
planarity of $\ell=4$ and $\ell=5$. 
We introduce the notion of a total likelihood to estimate the relevance of 
the low-multipoles tests of non-gaussianity. We show that, as a result of 
these tests, the CMB maps which are most widely used for cosmological
analysis lie within the $\sim$ 10\% of randomly generated maps
with lowest likelihoods.

\end{abstract}


\maketitle

\section{Introduction}

The cosmic microwave background (CMB) anisotropies have been measured 
with exquisite accuracy by WMAP, and the impact on Cosmology has been 
profound \cite{WMAP1y,dspergel,ILC1y,WMAP3y}. 
However, as the $\Lambda$CDM cosmological model becomes standard lore 
and the parameter space becomes narrower, the focus naturally drifts 
to the apparent anomalies. 
Among sources of concern that have survived the WMAP 3-year data are 
the lack of large-angle correlations \cite{WMAP1y,dspergel,WMAP3y}, 
which is mainly due to the low value of the cosmic quadrupole 
\cite{WMAP3y,OTZH,Efstathiou,Wagg,Prunet},
and the alignment between the quadrupole ($\ell=2$) and octopole 
($\ell=3$) \cite{WMAP3y,TOH,OTZH,Copi04,Schwarz04,KW04,jweeks,Eriksen04a,OT06}.

The combined statistics of these effects, which were already in
the COBE data \cite{COBE,cobe2}, implies that the probability that
our CMB sky was generated in a random process is only 0.005 -
0.02\%, depending on the map that is being tested.

Several recent works have reported some anomalies in the data: the
low-order multipole values~\cite{WMAP1y,dspergel,WMAP3y,Efstathiou,
TOH,Gaztanaga,Slosar,Copi06}; 
the alignment of some low-order 
multipoles~\cite{TOH,OTZH,Bielewicz04,Bielewicz05,LM05}; 
an unexpected asymmetric distribution on the sky of the large-scale 
power of CMB data~\cite{Hansen04a,Eriksen04b,Schwarz04,Copi05,BVWLF}; 
indications for a preferred direction of maximum asymmetry 
\cite{Copi04,Eriksen04b,Copi05,LM05b,BMRT,Wiaux}; 
as well as apparent non-gaussian features detected via the wavelet 
method or other 
analyses \cite{Vielva04,Cruz05,Cruz06,Hansen04b,McEwen06,BTV}. 
These anomalies have motivated many
explanations, such as compact topologies \cite{Luminet,MGRT04}, a
broken or suppressed spectrum at large scales
\cite{Broken,jcline,bfeng,stsuj,yspiao,Kawasaki}, oscillations 
superimposed on the primordial spectrum of density fluctuations
\cite{Jerome,jerome2,jerome3}, anisotropic cosmological models
\cite{Jaffe05,Cayon06,Gosh06} and possible extended foregrounds that
could be affecting the CMB \cite{Eriksen04a,AS,ReesSciama,IS06,ASW06}.

In this article we examine the multipoles $\ell=2-5$ and search
for anomalies in their phase correlations. We measure these
correlations through alignments between the multipole vectors 
or through their associated normal planes \cite{Copi04}. We 
also look for evidence
of  planarity (or {\em self-alignments}) in each individual
multipole, and for evidence of alignments between the multipole
and normal vectors with some specific directions in the sky, such
as the dipole axis, the ecliptic axis and the Galactic poles axis.
In total we have considered 38 tests of alignments between
multipoles, plus 8 tests of alignments of the multipoles with {\it
a priori} directions. We have computed the probability distribution
functions (PDF´s) for those tests using 300,000 mock maps.

For the statistical analyses peformed here, we need the $a_{\ell
m}$'s ($\ell=2-5$) of each CMB map under investigation. The
corresponding $a_{\ell m}$'s were extracted, after applying the
Kp2 WMAP mask (to minimize foreground contaminations from the
beginning), through the HEALPix routines \cite{Healpix}. Our
statistical tools were then used to analyze the WMAP 1-year and
3-year data Internal Linear Combination maps \cite{WMAP3y,WMAP1y}
(henceforth ILC), the co-added 1-year and 3-year WMAP data, as
well as the {\em cleaned} CMB maps of Tegmark {\it et} al.
\cite{TOH,OT06} based on 1-year WMAP data (henceforth TOH).

This paper is organized as follows. In Section II we summarize the
multipole vector formalism and the several different statistics
that can be used to test for alignments and phase correlations
within a given multipole. In Section III we briefly describe the
CMB maps used. Section IV presents our statistical tests and the
results of the PDF computations for those tests. We also analyze
the salient features of the CMB maps and discuss which tests can
be considered anomalous, and of those, which are robust and which
are most sensitive to noise. The conclusions are presented in
Section V.

\section{Multipole vectors, normal vectors and statistics of phase
correlations}

Multipole vectors were introduced in CMB data analysis by Copi 
{\it et al.} \cite{Copi04}, and Katz and Weeks~\cite{KW04,jweeks} 
found an elegant algebraic 
method to compute the multipole vectors given the spherical harmonic 
components $a_{\ell m}$ -- see also \cite{Germans} for an 
alternative algebraic method and 
\cite{OTZH,Eriksen04a,Schwarz04} for purely numerical methods. The
multipole vectors are essentially eigenvectors -- i.e., they
are solutions of a set of polinomial equations whose parameters 
are the multipole components $a_{\ell m}$.

The idea of the multipole vector representation goes back to J. C.
Maxwell in the XIX$^{\rm th}$ century: the multipole decomposition
of a field $f(\theta,\phi)$ on $S^2$ implies that for each multipole
$\ell$ there are $\ell$ eigenvectors of norm unity,
$\hat{n}^{(\ell,p)}$. Since there are only $2 \ell$ phases for
each multipole, the spherical harmonic representation and
the multipole vector representation have the same number of
degrees of freedom in each individual multipole:

\be
\label{expansions} \frac{\Delta T_\ell (\theta,\varphi)}{T} =
\sum_{m=-\ell}^\ell a_{\ell m} Y_{\ell m}
(\theta,\varphi) = D_\ell \prod_{p=1}^\ell
\hat{n}^{(\ell,p)} \cdot \hat{n} (\theta,\phi) - Z_{\ell-1} (\theta,\varphi) \; ,
\ee
where $Z_{\ell-1}$ just subtracts the residual $\ell ' < \ell$ total
angular momentum parts of the product expansion, and is irrelevant to
our analysis -- see \cite{KW04} for an enhanced discussion of the
multipole vector expansion.

It can be seen from the product expansion above that, whenever using 
the multipole vectors to test for alignments, it is irrelevant what
the amplitudes of the multipoles are -- just their phases matter.
This is the main feature of the tests based on the multipole
vectors that sets them apart from other tests of non-gaussianity.

Notice that, contrary to the $C_\ell$'s, which are always
positive-definite, the $D_\ell$'s of Eq. (\ref{expansions}) can
be either negative or positive. Because of the product expansion
in the right-hand-side of Eq. (\ref{expansions}), switching the
sign of $D_\ell$ is equivalent to switching the signs of an odd
number of multipole vectors, and switching the signs of an even
number of multipole vectors leaves the sign of $D_\ell$ invariant.
Therefore, the product expansion in Eq. (\ref{expansions}) has a
sign degeneracy in the amplitudes $D_\ell$ as well as in the the
multipole vectors $\hat{n}^{(\ell,p)}$.

This ``reflection symmetry" $\hat{n}^{(\ell,p)} \leftrightarrow -
\hat{n}^{(\ell,p)}$ implies that the multipole vectors define only
directions \cite{KW04}, hence they are ``vectors without
arrowheads" living on the half-sphere with antipodal points
identified, or $S^2/{\mathbb Z}_2$. This space is also known in
the literature as the real projective space ${\mathbb R} P^2$, and
is useful in the characterization of nematic liquid crystals,
where the orientation of the molecules is an order parameter --
but it makes no difference where heads and tails are
\cite{Nakahara}.

It is extremely useful to represent these directions as vectors,
but for that we will need to cope with the degeneracy in
representing these directions. For each multipole order $\ell$
there is a $2^{\ell-1}$-fold degeneracy in the signs (or
orientations) of the multipole vectors -- corresponding to the
$\ell$ signs of the multipole vectors that can be switched
arbitrarily, divided by two to account for an irrelevant overall
sign which is determined by $D_\ell$. We can break this degeneracy
by always working in one particular hemisphere, and any such
choice will automatically determine the signs (orientations) of
all multipole vectors -- as well as the sign of the $D_\ell$'s.
However, we should be aware that in doing so we are necessarily
picking {\it one} of the $2^{\ell-1}$ possible sign conventions in
the product expansion of Eq. (\ref{expansions}). As we will
discuss below, this is not a problem as long as we use invariant
tools which are not sensitive to the signs of each multipole
vector.











Starting with the $\ell$ multipole vectors one can also construct
$\ell(\ell-1)/2 \equiv \lambda$ normal vectors -- or normal planes
-- defined as:
\be \label{normals} \vec{w}^{\ell,q} \equiv \hat{n}^{\ell,p}
\wedge \hat{n}^{\ell,p'} \quad , \quad (p\neq p' \; , \; q = 1 \ldots \lambda) \; .
\ee

By construction, because the multipole vectors define only
directions, the normal vectors also possess reflection symmetry,
$\vec{w}^{(\ell,q)} \leftrightarrow - \vec{w}^{(\ell,q)}$. But the
normal vectors need not be (and generally are not) of norm unity,
so instead of living in $S^2/{\mathbb Z}_2$ the normal vectors
belong to the space ${\mathbb R}^3/{\mathbb Z}_2$ -- which is 
isomorphic to SO(3), see \cite{Nakahara}. We can still
break the degeneracies imposed by reflection symmetry by defining
all normal vectors so they lie in the same hemisphere as the
normal vectors. However, just as before, we should be aware that
in doing so we are choosing one of many possible representations
for the normal vectors.

To summarize, for $\ell=2$ there are 2 multipole vectors,
$\hat{n}^{(2,1)}$ and $\hat{n}^{(2,2)}$, and only one normal
vector, $\vec{w}^{(2,1)}= \hat{n}^{(2,1)} \times \hat{n}^{(2,2)}$;
for $\ell=3$ there are 3 multipole vectors and 3 normal vectors;
and so forth. These constitute the basis for the statistical tests
defined below.

\subsection{Properties of the tests under reflection symmetry}

We will define below, in Sec. IIC, a series
of tests which are manifestly invariant
under the reflection symmetry $\hat{n}
\leftrightarrow -\hat{n}$ which characterizes 
${\mathbb R}P^2$ (where the multipole vectors live) as
well as ${\mathbb R}^3/{\mathbb Z}_2$ (where the associated
normal vectors live.)

Our motivation for this remark is that if a given test
is not invariant then its validity and usefulness is questionable. 
In particular, one should be careful
not to employ tests which depend on the choice of hemisphere to
represent the vectors. This seems to be the case of some of the
tests that have been used to estimate the ``planarity" of the CMB
maps, if these tests make use, in one way or another, of the
notion of ``average vectors". The reason there is no such thing as
an ``average multipole vector" or an ``average normal vector" is
simple: the ``vector sum" operation does not yield a singly valued
result due to the reflection symmetry. In fact, the result of
``summing" two vectors of ${\mathbb R}^3/{\mathbb Z}_2$ would be a
degenerated pair of directions: 
\be 
\label{sumvec} 
(\pm \vec{v}_1) \oplus (\pm \vec{v}_2) 
= \left\{
\begin{array}{c}
\pm \vec{v}_1 \pm \vec{v}_2 \\
\pm \vec{v}_1 \mp \vec{v}_2
\end{array}
\right. \; .
\ee

In general, by ``summing" $k$ vectors of ${\mathbb R}^3/{\mathbb Z}_2$
one obtains $2^{k-1}$ vectors, corresponding to the $2^{k}$ possible
permutations of the $\pm$ signs of each vector,
divided by two to account for the
symmetry $\vec{v} \leftrightarrow -\vec{v}$ of the resulting vector.
This means, in particular, that the ``average multipole vector" is in
fact an object $2^{\ell-1}$-times degenerated,
and that the ``average normal vector" is an object
$2^{\ell(\ell-1)/2-1}$-times degenerated.

Obviously, by fixing a hemisphere to represent all vectors one
breaks this degeneracy, but this just hides the plain fact that by
doing so one is simply choosing (rather arbitrarily) one of many
possible representations, and one of many possible answers for the
sums of those vectors. Consequently, unless these degeneracies are
properly taken into account (by, e.g., symmetrizing over all
possibilities or ordering the results by norm), any test which is 
derived from the notion of summing multipole or normal vectors 
is flawed.

\subsection{Global estimates of non-gaussianity}

We will investigate large-scale correlations in the CMB maps
within single multipoles and between different multipoles by
measuring the alignments between either the multipole vector
themselves, or between their associated normal planes. To be sure,
there is no upper limit to the number of tests
we can devise to search for
non-gaussianities, and in testing any fixed sample such as the CMB
one should bear in mind that there are
always some statistical tools which will yield a positive
detection given some arbitrary criteria.

In practical terms this means that if we perform a large number
${\cal{N}}$ of independent tests on a fixed sample, then we should
treat the results of these tests themselves as random numbers.
Therefore, when performing many tests on a map and searching for
clues of non-gaussianity one must always look at the complete set
of results for those tests and at the total probability (or
some estimate of the likelihood) that map is a realization of 
a random process. If
that likelihood turns out to be very small compared to the typical
likelihoods of gaussian maps, then one can look for
the particular test (or tests) that is likely responsible for that
anomaly. If, however, a certain test turns out to be ``suspicious"
but the likelihood is not anomalously small, then we cannot rule
out the possibility that the result for that particular test was
just a fluke.

Assuming that random processes are indeed behind the mechanism
that generated the sample, we can define a total likelihood in the
following sense. Suppose we have ${\cal{N}}$ statistical tests
such that each test $T_i$ ($i=1\ldots {\cal{N}}$) is a random
number in the interval $0 \leq T_i \leq 1$, with normalized
probability distribution functions $P_i(T_i)$. Given a sample (a
map $M$) with $T_i = T_i^M$, the probability that a random sample
has a value of $T_i$ {\it higher} than $T_i^M$ is:
\be
\label{P+} P_{i+}(T_i^M) = \int_{T_i^M}^1 dT P_i(T) \; ,
\ee
and the probability that a random sample has a value of $T_i$
{\it lower} than $T_i^M$ is:
\be
\label{P-} P_{i-}(T_i^M) = \int_{0}^{T_i^M} dT P_i(T) = 1-
P_{i+}(T_i^M) \; .
\ee
Evidently, for the median value $\bar{T}_i$ we have
$P_{i+}(\bar{T}_i)=P_{i-}(\bar{T}_i)=1/2$. We will {\it define} 
the likelihood of the map $M$, given the ${\cal{N}}$ tests, to be:
\be
\label{likelihood} L_{\cal{N}} (M) = \prod_{i=1}^{\cal{N}} 2
P_{i+}(T_i^M) \times 2 P_{i-}(T_i^M) = 4^{{\cal{N}}}
\prod_{i=1}^{\cal{N}}  P_{i+}(T_i^M) \left[ 1- P_{i+}(T_i^M)
\right] \; ,
\ee
where the factors of 2 have been inserted for normalization
purposes, in order to make $L_{\cal{N}}({\bar{M}})=1$ for a
map $\bar{M}$ whose tests are all exactly equal to their median
values. This likelihood measures the total probability that the
map $M$ does not have too low {\it and} too high values of the
tests $T_i$. Evidently, any deviation of the tests from the
medians will decrease $L_{\cal{N}}$.

Of course, Eq. (\ref{likelihood}) is itself an arbitrary
definition, and we might as well have used the expectation values
instead of the medians to define the likelihood. Indeed,
one could switch the roles of the medians
with the expectation values in the procedure above and still our
results would be very similar.

It should be stressed that the total likelihood defined in the 
sense above should {\it not} be interpreted as the probability that
a particular map was generated by a gaussian mechanism -- it 
is merely an estimator of how much that particular map deviates 
from a typical one, given the ${\cal{N}}$ statistical tests of 
non-gaussianity.
In Sec. IID we construct such likelihoods, and compute their
distributions assuming random phases.

\subsection{Statistical tests of isotropy}

We now define the statistical tests which will be employed in our
analysis of CMB data. We have ensured that all tests are invariant
under reflection symmetry, so it makes no difference which
hemisphere one chooses to represent the multipole and normal
vectors.

The tests have been normalized so that they always fall in the
interval $0\leq T_i \leq 1$. We have generated $3 \times 10^5$ 
simulated (mock) maps using gaussian random phases, and the 
resulting PDF's for the tests are shown in Figs. 1-5.

\subsubsection{$S$ statistic}

The $S$ statistic is a widely used tool \cite{Schwarz04,KW04}, and it measures the
alignment between normal planes of different multipoles. It is defined as:
\be
\label{def:S}
S_{\ell \ell'} \equiv \frac{1}{\lambda \lambda'}
\sum_{q=1}^\lambda \sum_{q'=1}^{\lambda'}
\left| \vec{w}^{(\ell,q)} \cdot \vec{w}^{(\ell',q')} \right| \quad , \quad \ell\neq\ell' \; ,
\ee
where, as defined above, $\lambda=\ell(\ell-1)/2$.
We can also use $S$ in just one multipole, in which case the normalization
is a bit different:
\be
\label{def:S2}
S_{\ell \ell} \equiv \frac{2}{\lambda (\lambda-1)}
\sum_{q,q'>q}^\lambda
\left| \vec{w}^{(\ell,q)} \cdot \vec{w}^{(\ell,q')} \right| \; .
\ee
The statistic $S_{\ell \ell}$ measures the ``self-alignment" of
the normal vectors, and is related to the ``planarity" tests
\cite{OTZH,Schwarz04}.

\begin{figure}[t]
\includegraphics[width=5.5cm]{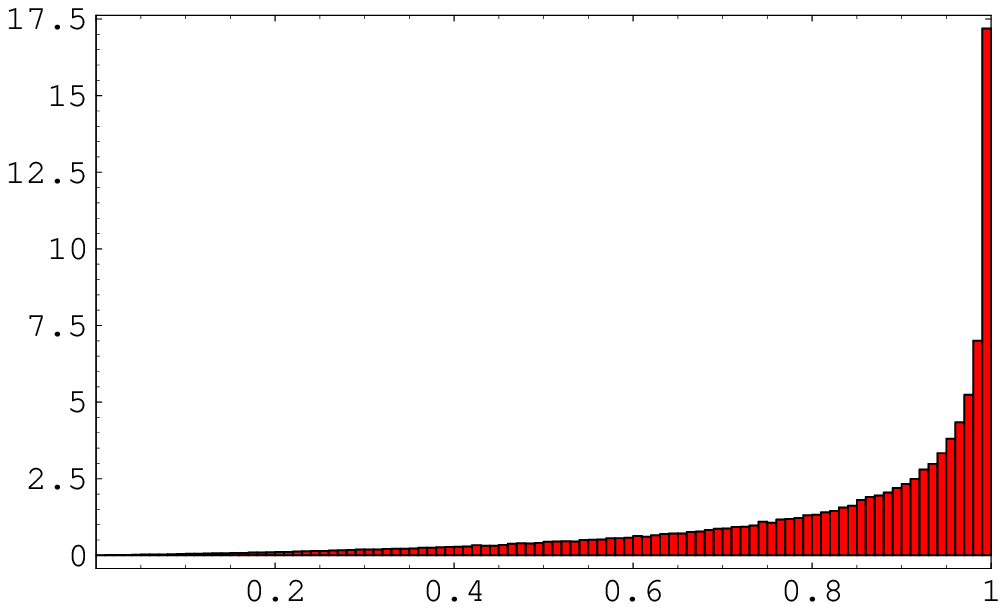}
\includegraphics[width=5.5cm]{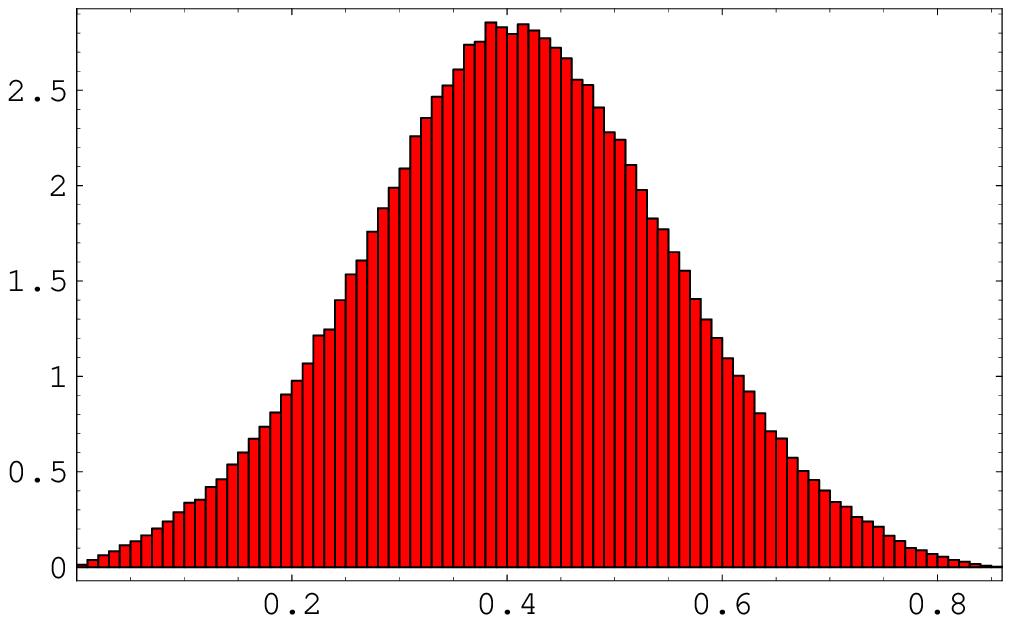}
\includegraphics[width=5.5cm]{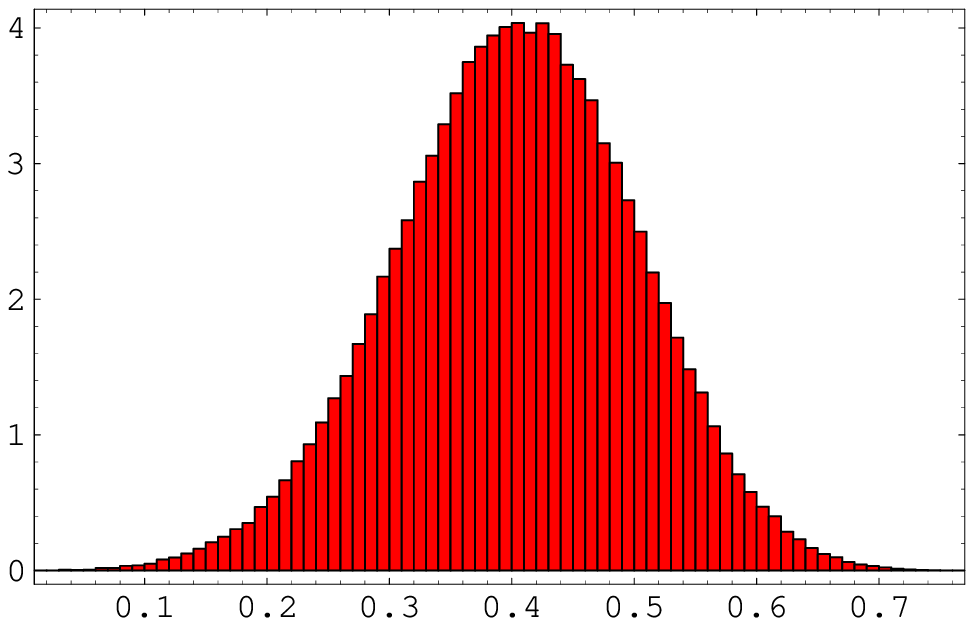}
\includegraphics[width=5.5cm]{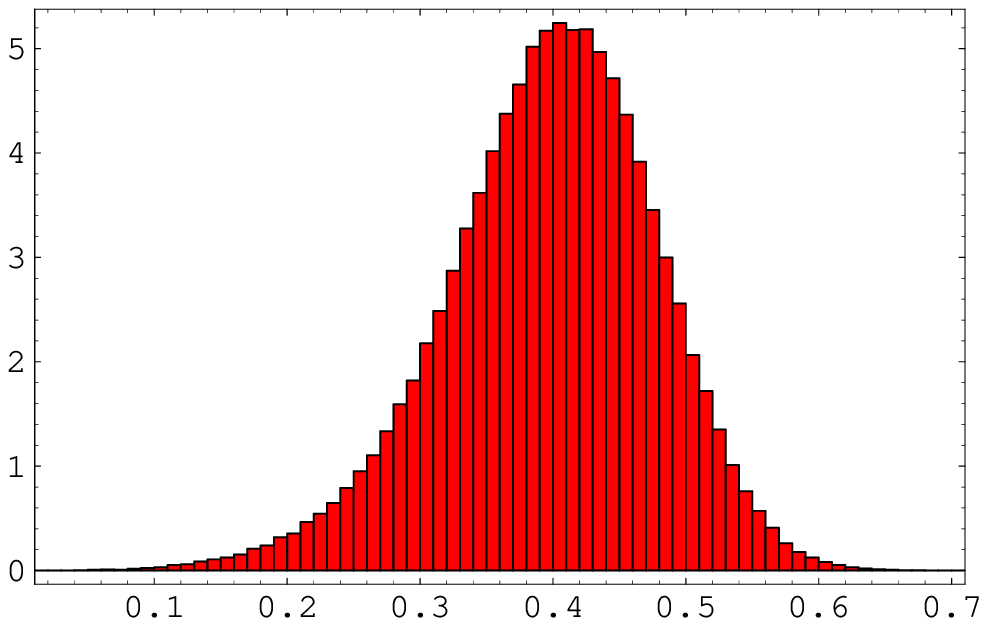}
\includegraphics[width=5.5cm]{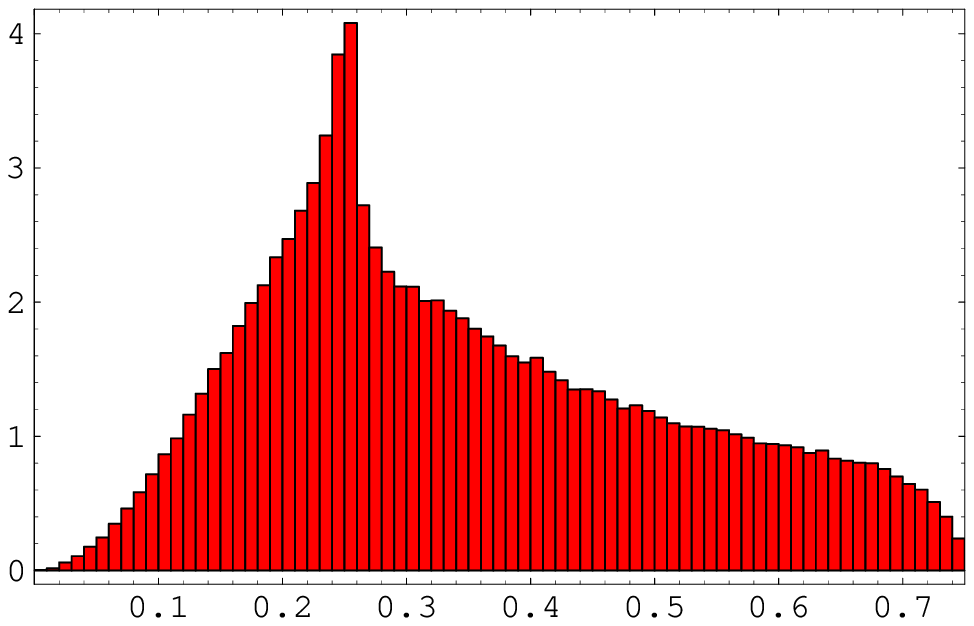}
\includegraphics[width=5.5cm]{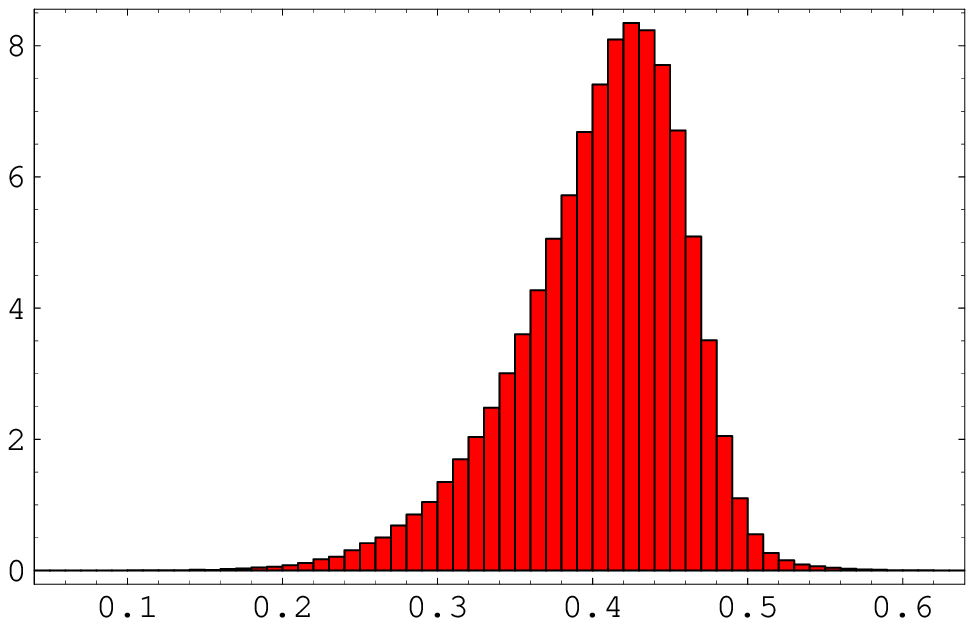}
\includegraphics[width=5.5cm]{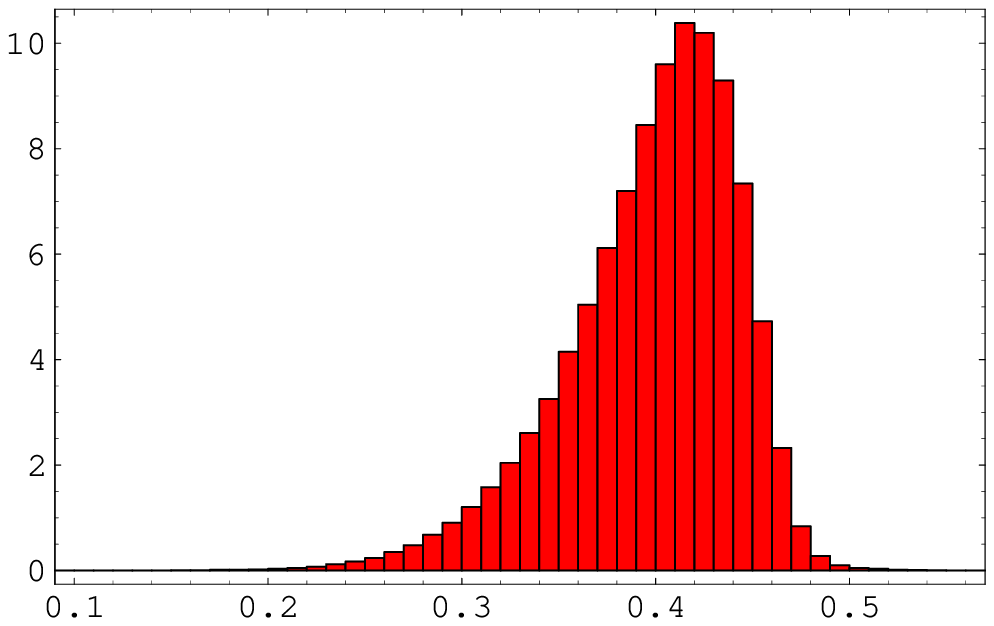}
\includegraphics[width=5.5cm]{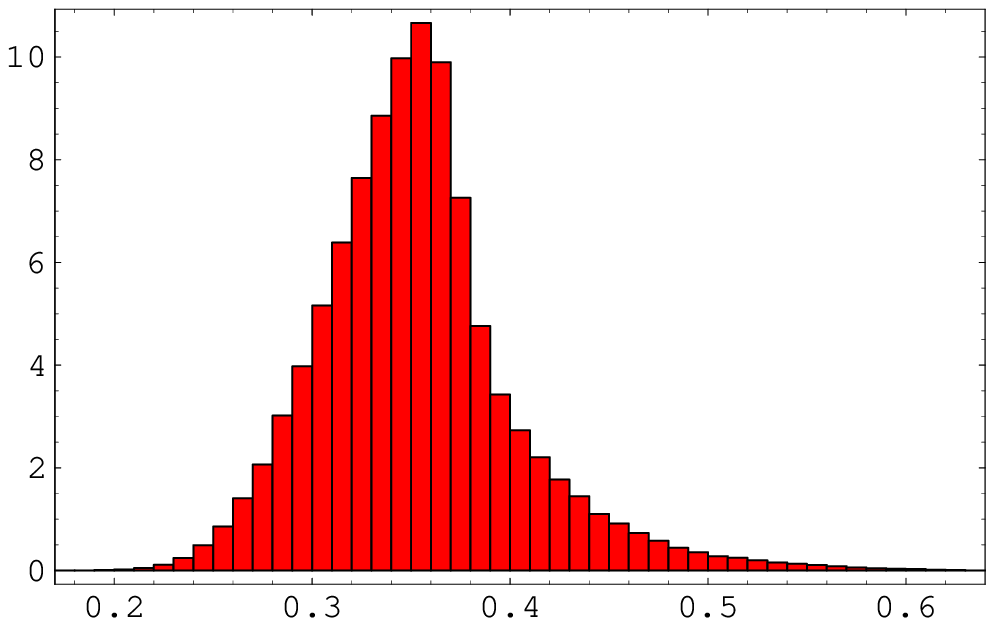}
\includegraphics[width=5.5cm]{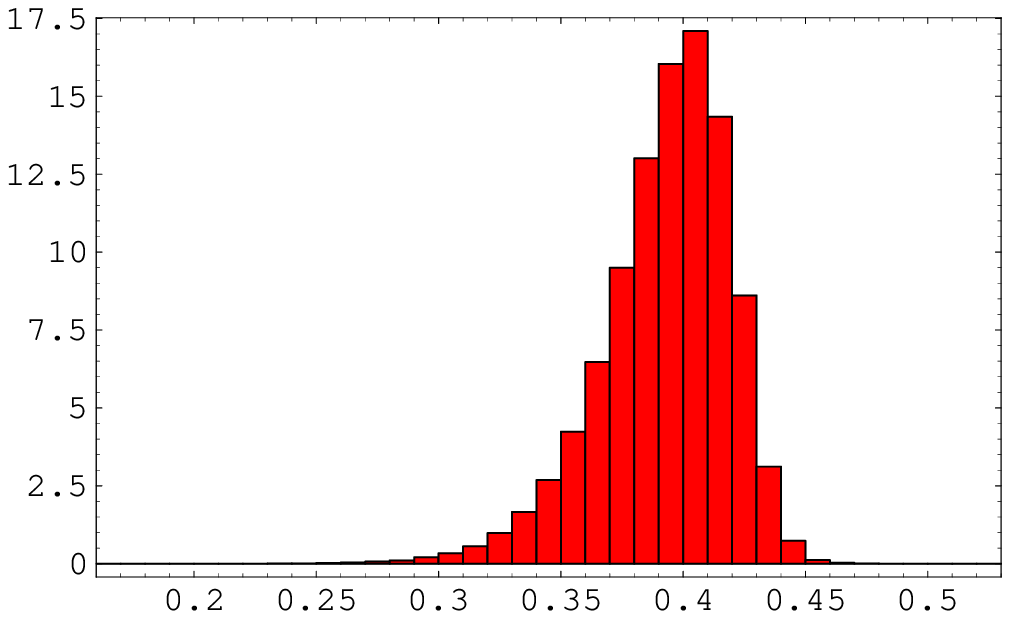}
\includegraphics[width=5.5cm]{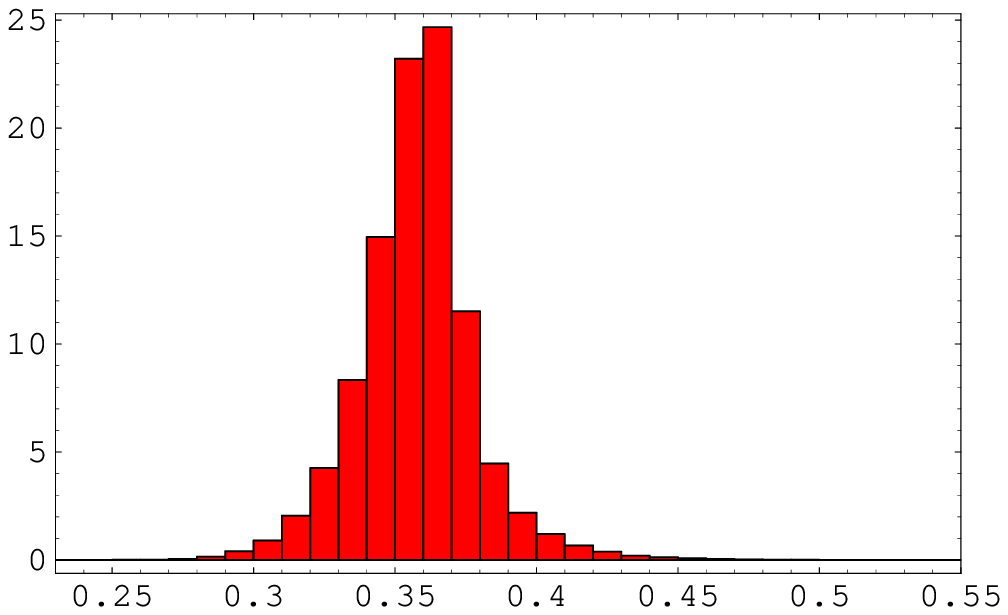}
\caption{\label{fig:histograms1} Normalized PDF's for the $S$
statistic found by simulating $3\times 10^5$ mock maps, binned
in intervals of 0.01. 
From left to right, top to bottom: $S_{22}$, $S_{23}$,
$S_{24}$, $S_{25}$, $S_{33}$, $S_{34}$, $S_{35}$, $S_{44}$,
$S_{45}$ and $S_{55}$. In all panels, the horizontal axis correspond to the value
of each individual test, and the vertical axis to its normalized
PDF.}
\end{figure}

\subsubsection{$D$ statistic}

This is analogous to the $S$ statistic, but the $D$ test
disregards the norm of the normal vectors: 
\be 
\label{def:D}
D_{\ell \ell'} \equiv \frac{1}{\lambda \lambda'}
\sum_{q=1}^\lambda \sum_{q'=1}^{\lambda'} \left|
\hat{w}^{(\ell,q)} \cdot \hat{w}^{(\ell',q')} \right| \quad ,
\quad \ell\neq\ell' \; . 
\ee 
We can also use the $D$ statistic
within a single multipole, as was done for $S$. However, this test
only gives nontrivial information for $\ell \geq 3$. With the
proper normalization we have:
\be
\label{def:D2}
D_{\ell \ell} \equiv \frac{2}{\lambda
(\lambda-1)} \sum_{q,q'>q}^\lambda \left| \hat{w}^{(\ell,q)} \cdot
\hat{w}^{(\ell,q')} \right| \quad , \quad \ell \geq 3 \; .
\ee

\begin{figure}[t]

\includegraphics[width=5.5cm]{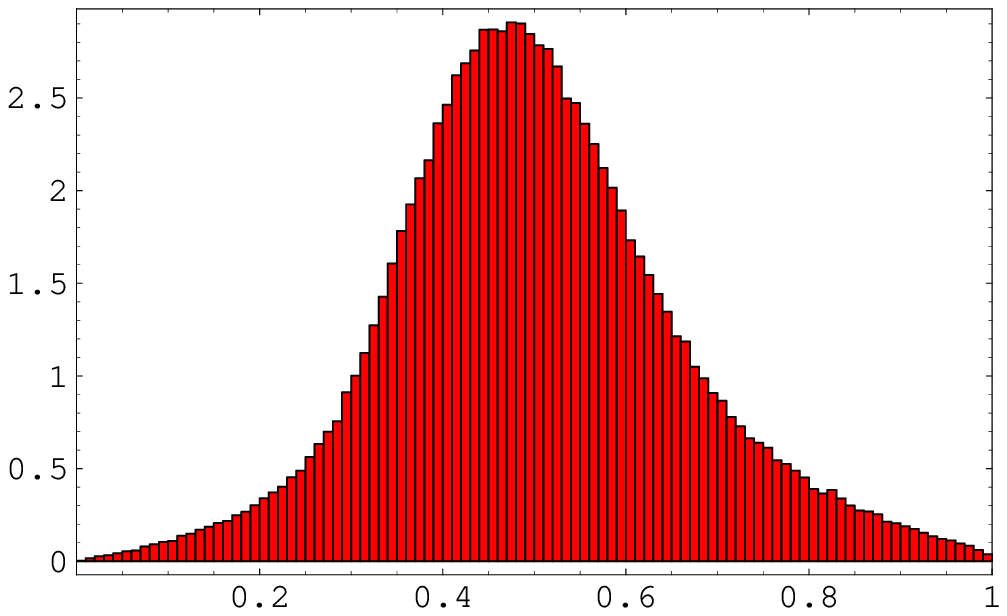}
\includegraphics[width=5.5cm]{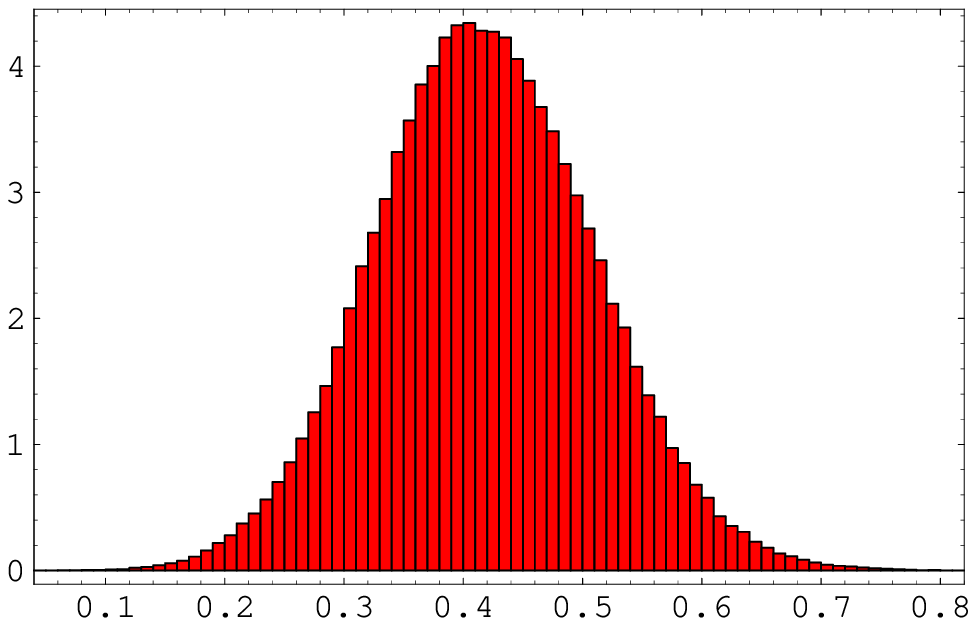}
\includegraphics[width=5.5cm]{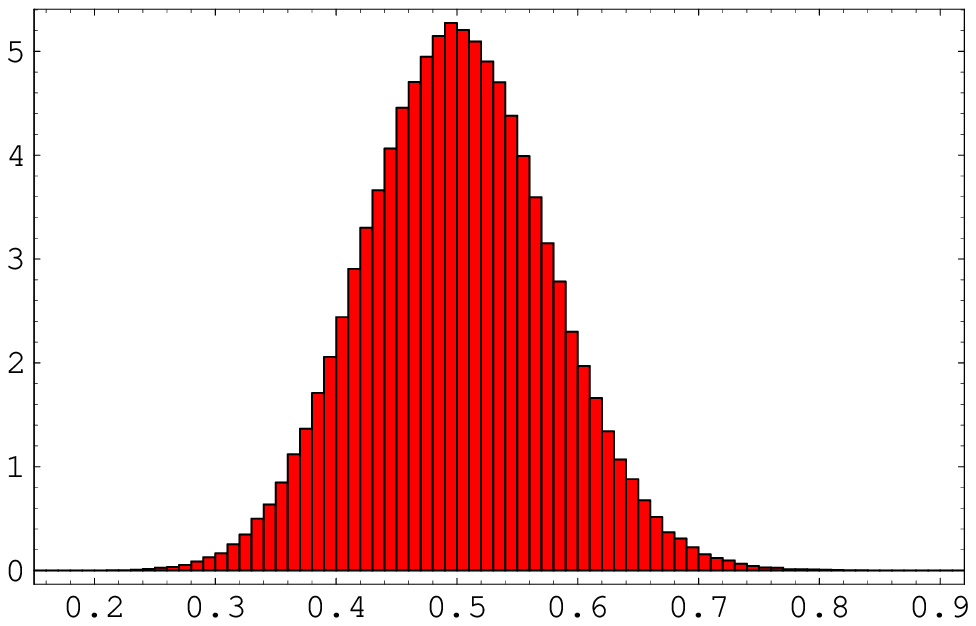}
\includegraphics[width=5.5cm]{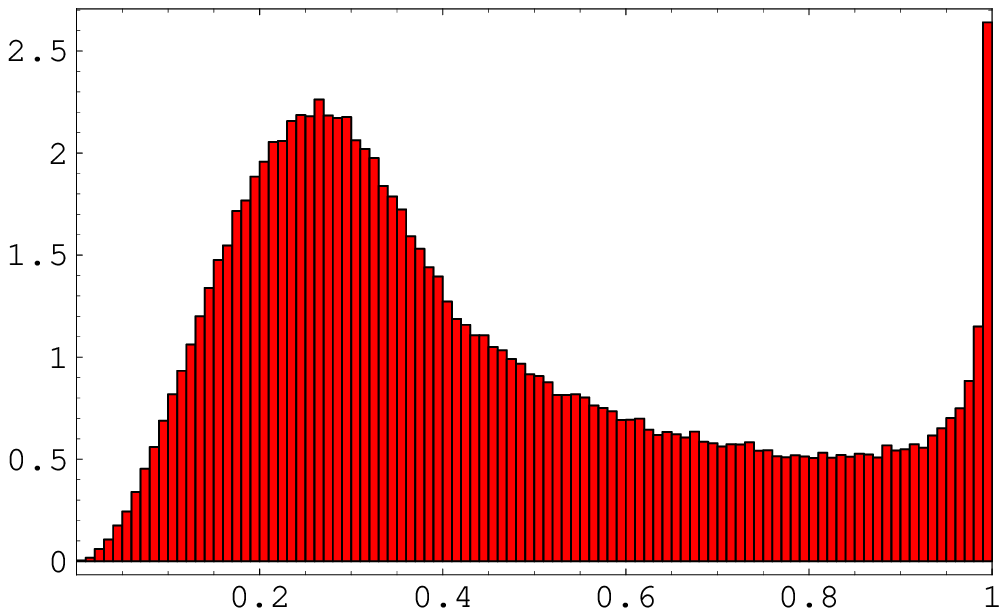}
\includegraphics[width=5.5cm]{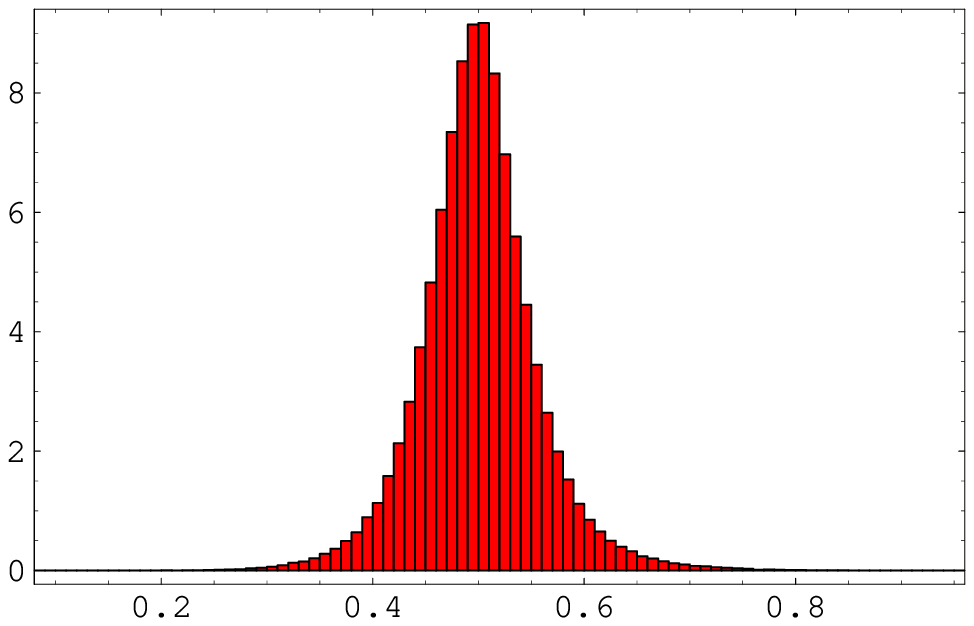}
\includegraphics[width=5.5cm]{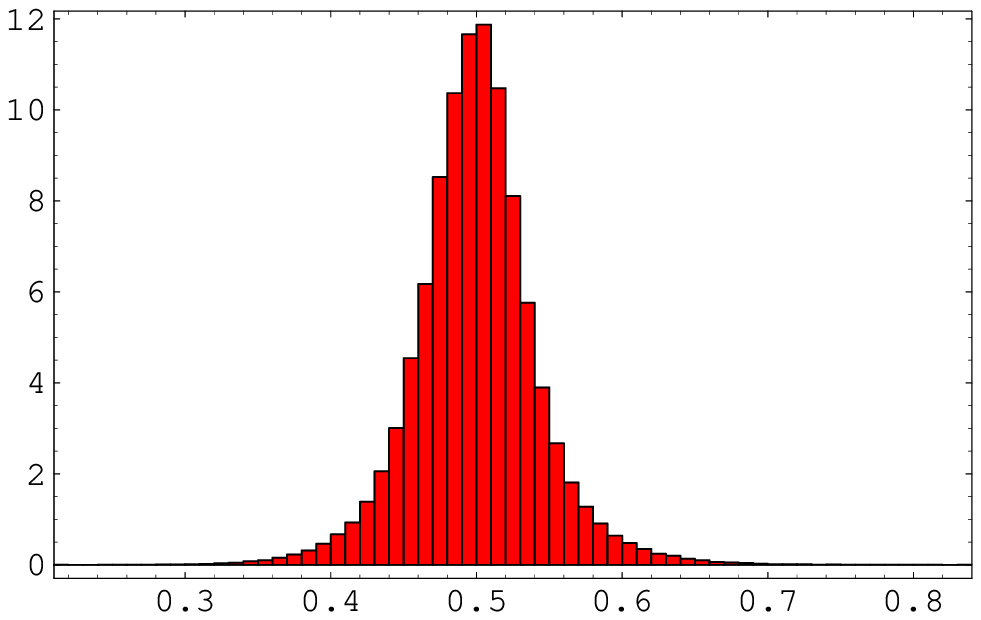}
\includegraphics[width=5.5cm]{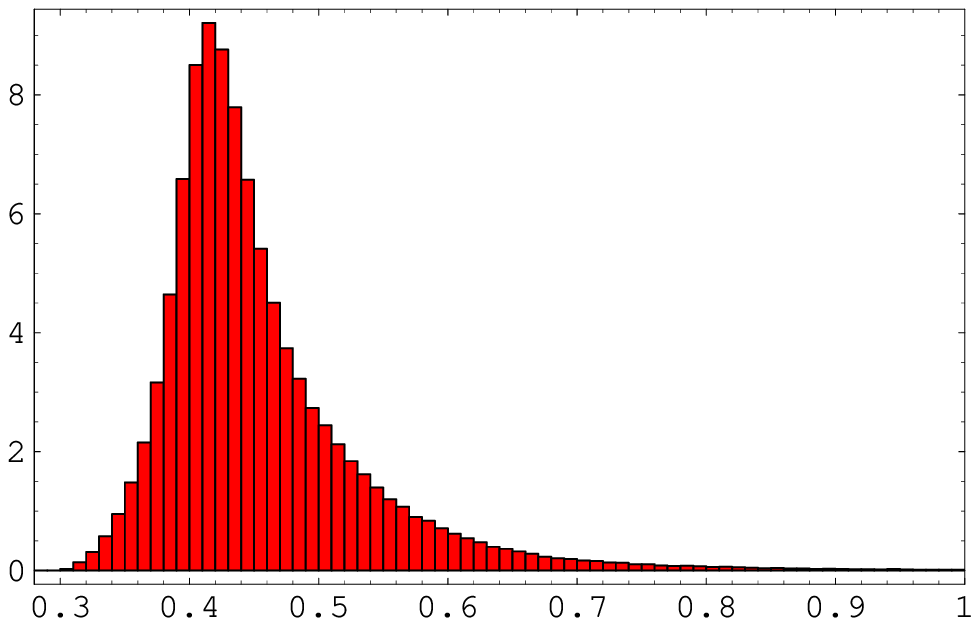}
\includegraphics[width=5.5cm]{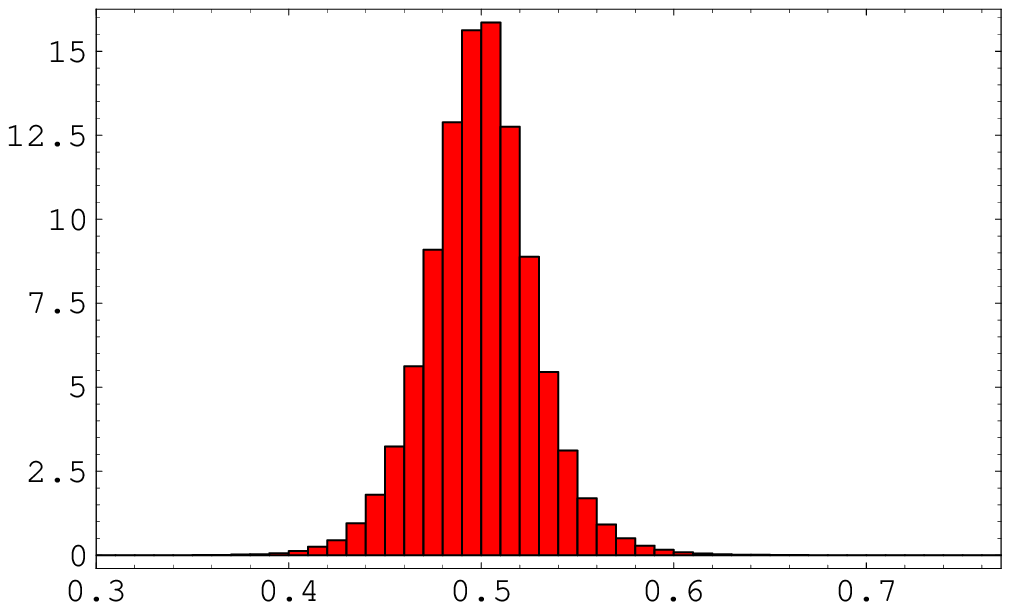}
\includegraphics[width=5.5cm]{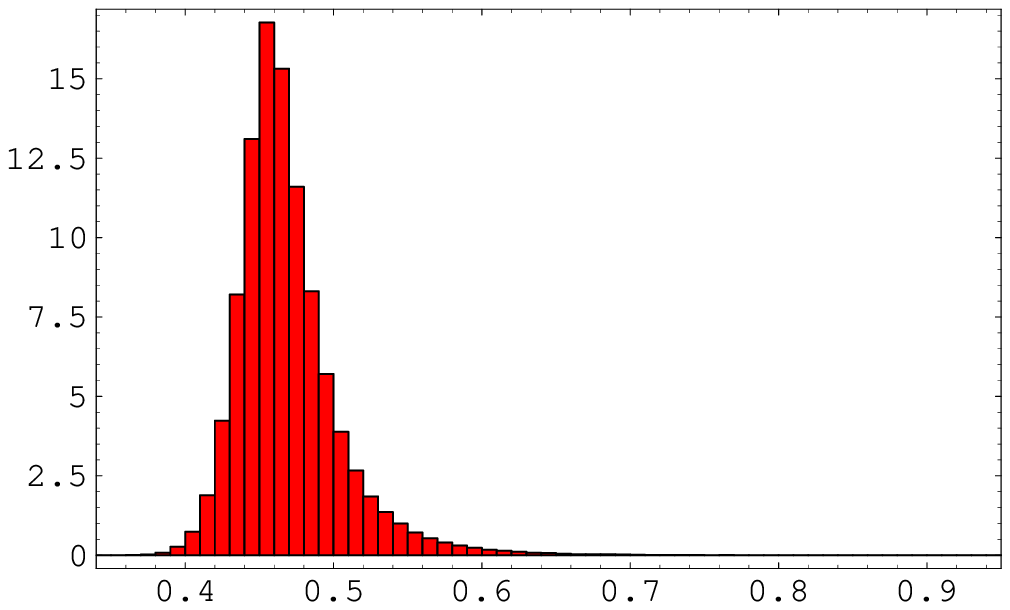}

\caption{\label{fig:histograms2} Normalized PDF's for the $D$
statistic. From left to right, top to bottom: $D_{23}$, $D_{24}$,
$D_{25}$, $D_{33}$, $D_{34}$, $D_{35}$, $D_{44}$, $D_{45}$ and
$D_{55}$.}

\end{figure}

\subsubsection{$R$ statistic}

A similar tool is the $R$ statistic, which measures alignments
in essentially the same way as the $S$ statistic, but it uses
the multipole vectors instead of the normal vectors:
\be
\label{def:R}
R_{\ell \ell'} \equiv \frac{1}{\ell \ell'}
\sum_{p=1}^\ell \sum_{p'=1}^{\ell'}
\left| \hat{n}^{(\ell,p)} \cdot \hat{n}^{(\ell',p')} \right| \quad , \quad \ell\neq\ell' \; .
\ee
Within a single multipole, the $R$ statistic is suitably defined as:
\be
\label{def:R2}
R_{\ell \ell} \equiv \frac{2}{\ell (\ell-1)}
\sum_{p,p'>p}^\ell
\left| \hat{n}^{(\ell,p)} \cdot \hat{n}^{(\ell,p')} \right| \; .
\ee

\begin{figure}[t]

\includegraphics[width=5.5cm]{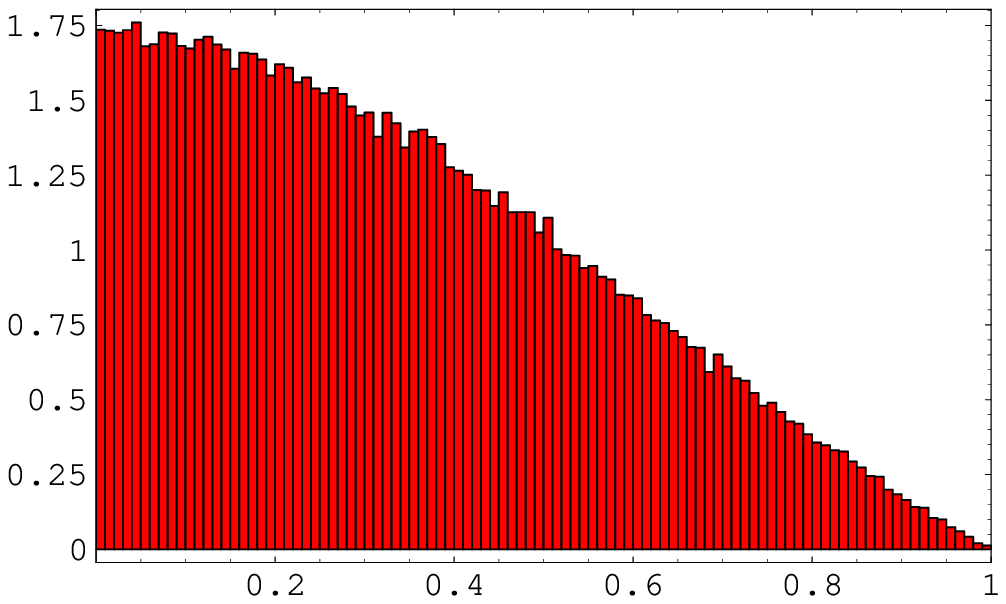}
\includegraphics[width=5.5cm]{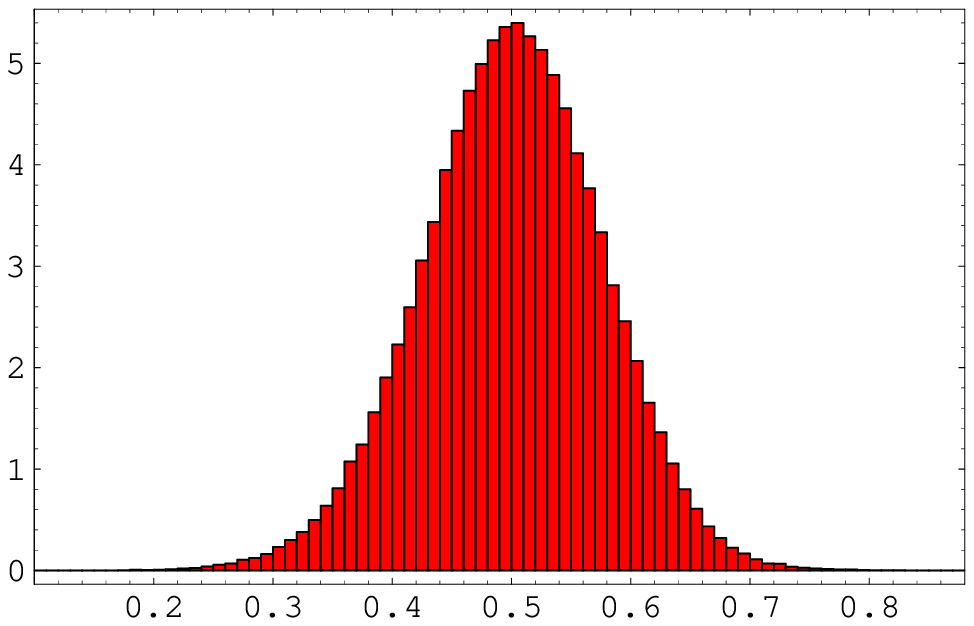}
\includegraphics[width=5.5cm]{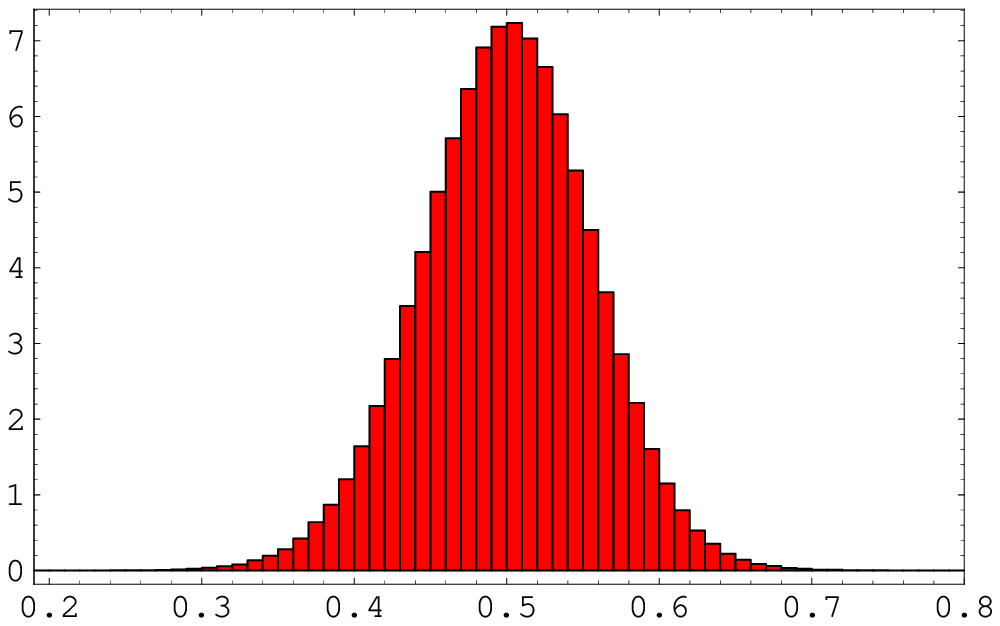}
\includegraphics[width=5.5cm]{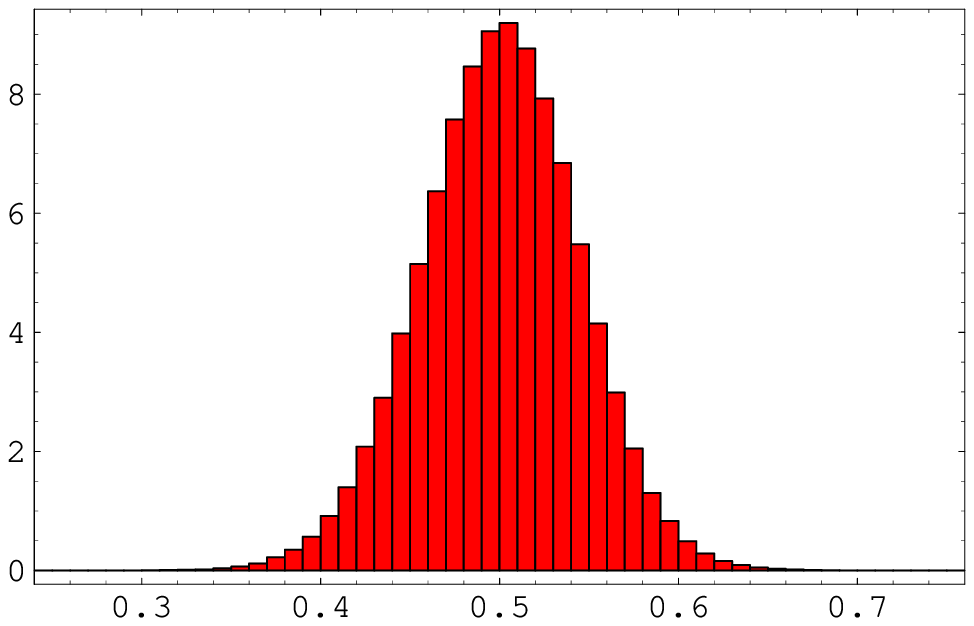}
\includegraphics[width=5.5cm]{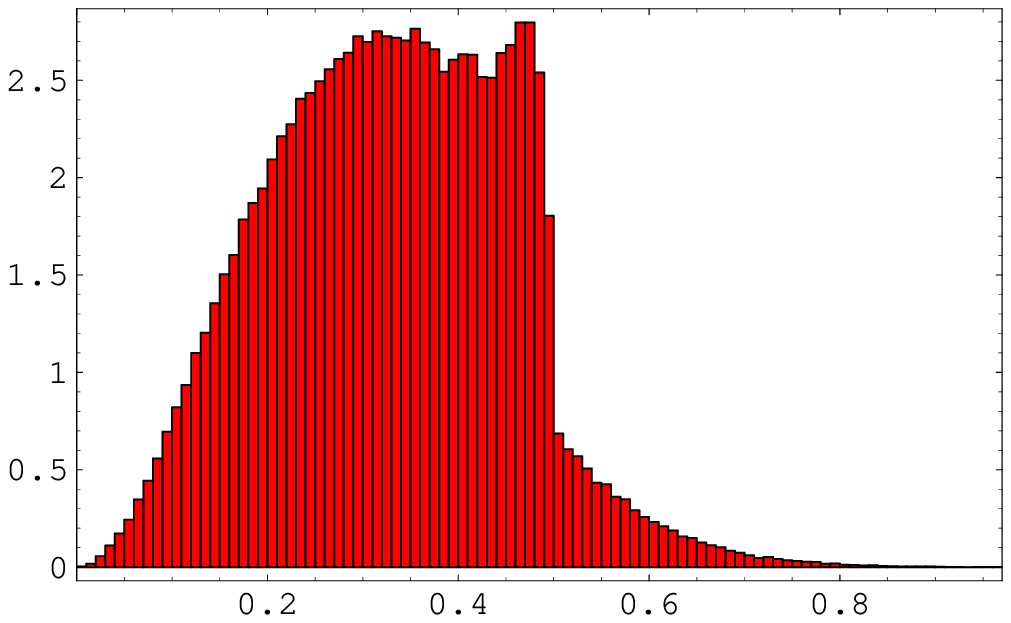}
\includegraphics[width=5.5cm]{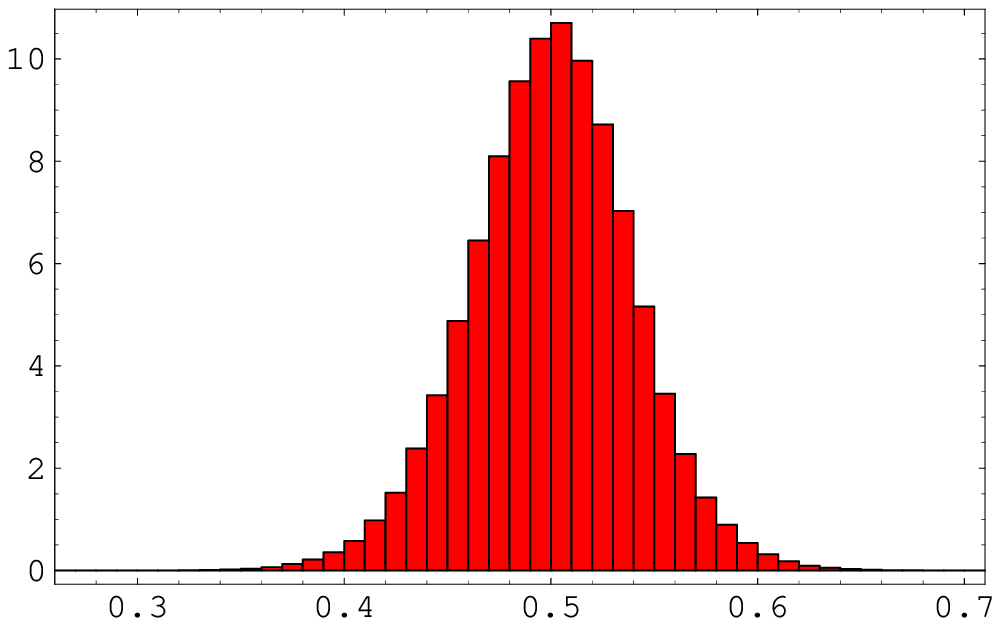}
\includegraphics[width=5.5cm]{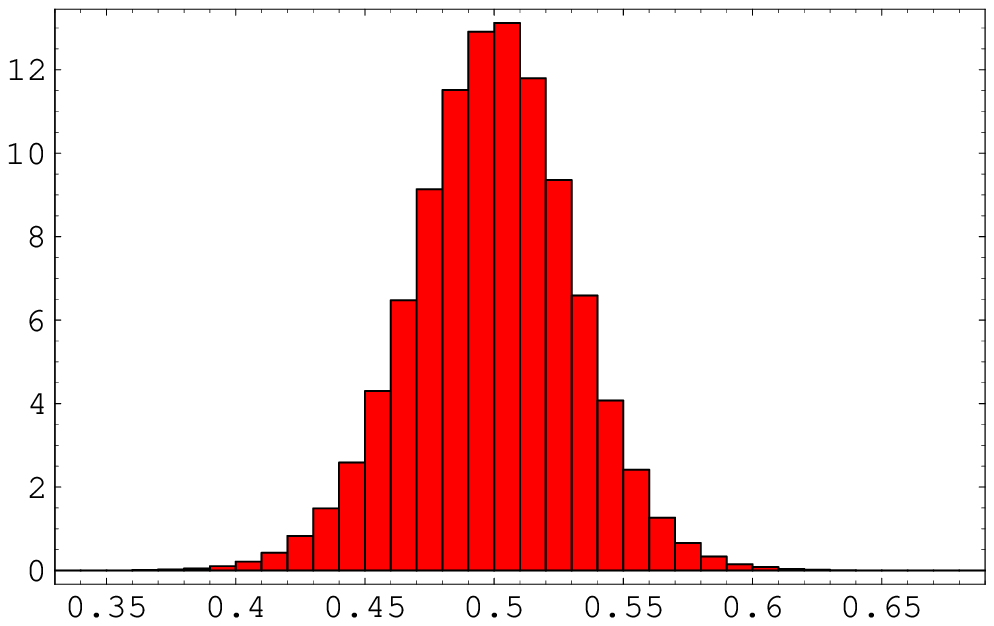}
\includegraphics[width=5.5cm]{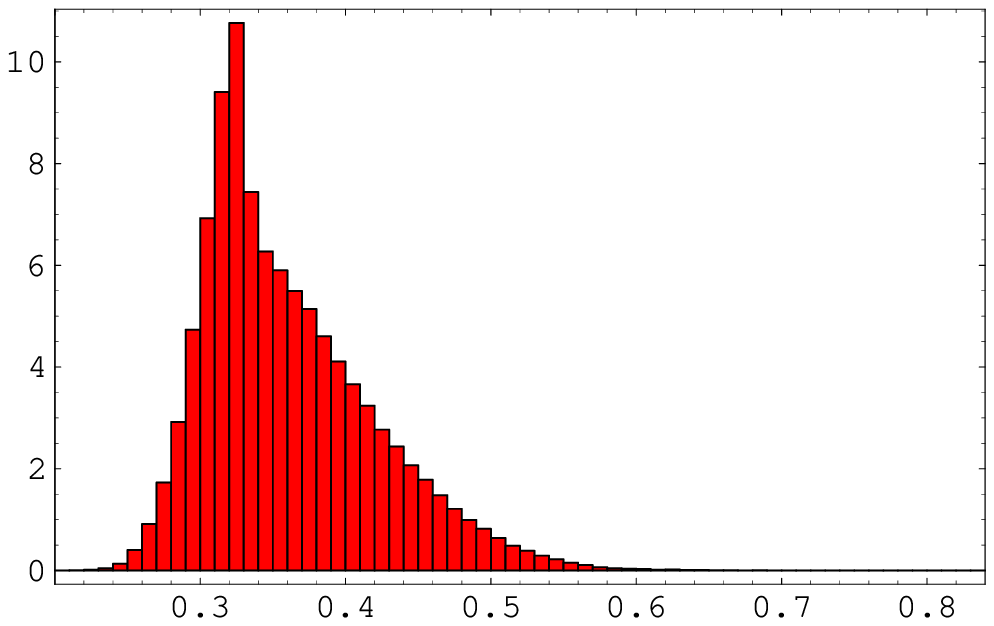}
\includegraphics[width=5.5cm]{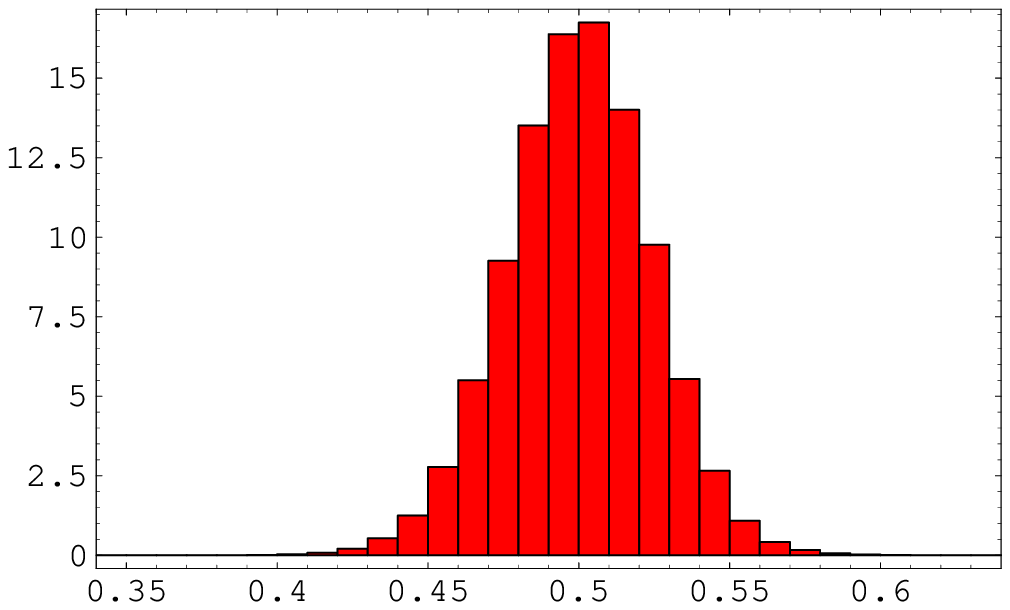}
\includegraphics[width=5.5cm]{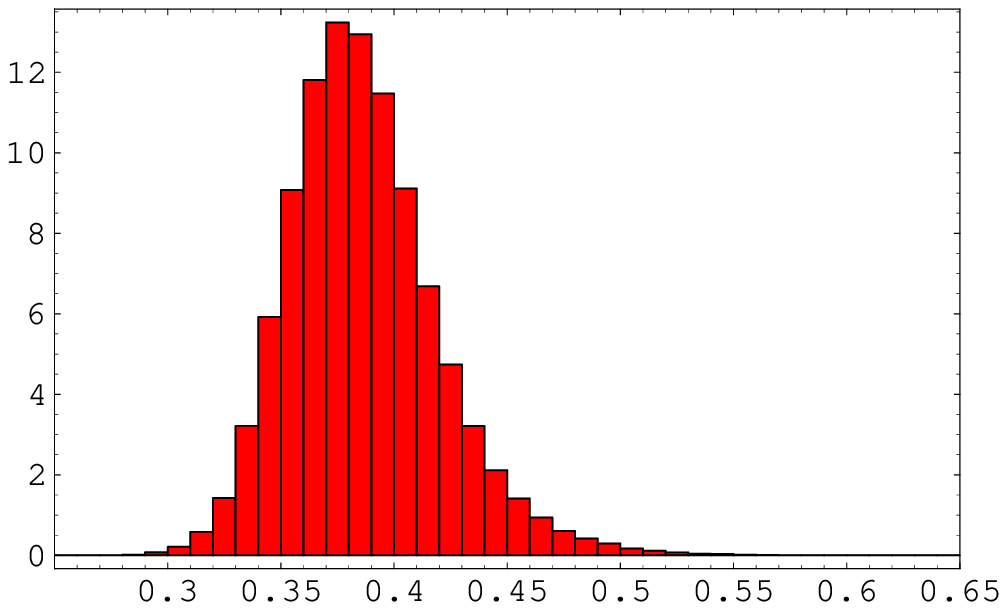}

\caption{\label{fig:histograms3} Normalized PDF's for the $R$
statistic. From left to right, top to bottom: $R_{22}$, $R_{23}$,
$R_{24}$, $R_{25}$, $R_{33}$, $R_{34}$, $R_{35}$, $R_{44}$,
$R_{45}$ and $R_{55}$.}

\end{figure}

\subsubsection{$B$ statistic}

We can also test if the multipole vectors align with the normal
vectors. Hence we define the $B$ statistic:
\be
\label{def:B}
B_{\ell \ell'} \equiv \frac{1}{\lambda \ell'}
\sum_{q=1}^\lambda \sum_{p'=1}^{\ell'}
\left| \vec{w}^{(\ell,q)} \cdot \hat{n}^{(\ell',p')} \right| \quad , \quad \ell\neq\ell' \; .
\ee
Within a single multipole, the $B$ statistic only gives nontrivial
information for $\ell \geq 3$, and we have:
\be
\label{def:B2}
B_{\ell \ell} \equiv \frac{1}{\lambda (\ell-2)}
\sum_{q}^\lambda \sum_{p}^\ell
\left| \vec{w}^{(\ell,q)} \cdot \hat{n}^{(\ell,p)} \right| \quad , \quad \ell \geq 3 \; .
\ee
Notice that, as opposed to the $S$, $D$ and $R$ statistics, the $B$ statistic
for $\ell \neq \ell'$ is not symmetric, $B_{\ell \ell'} \neq B_{\ell' \ell}$.
For simplicity, in the present approach we have only considered the cases
$B_{\ell \ell'}$ where $\ell \leq \ell'$.

\begin{figure}[t]

\includegraphics[width=5.5cm]{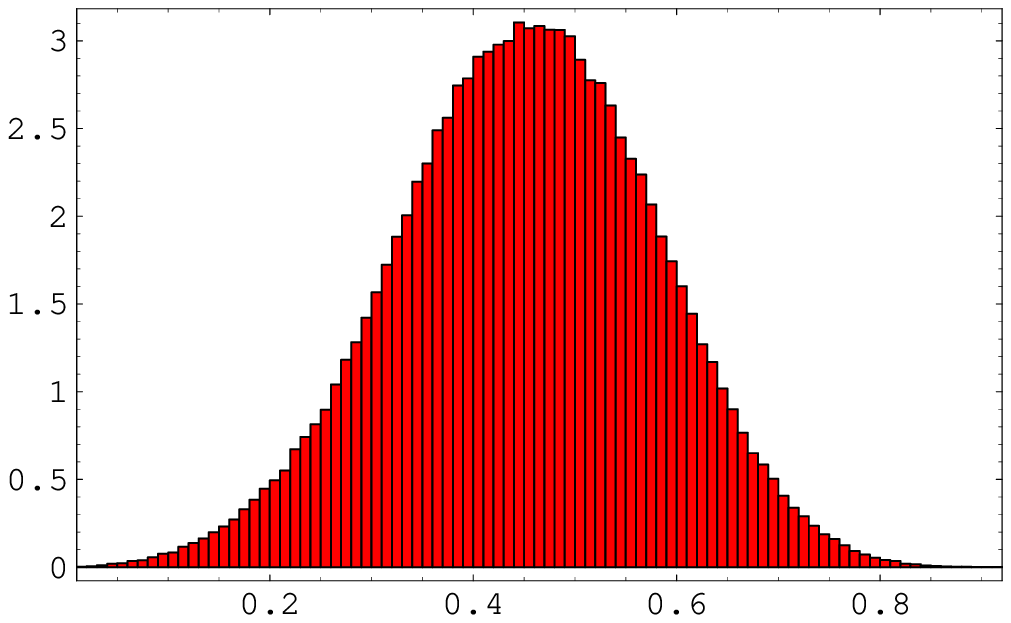}
\includegraphics[width=5.5cm]{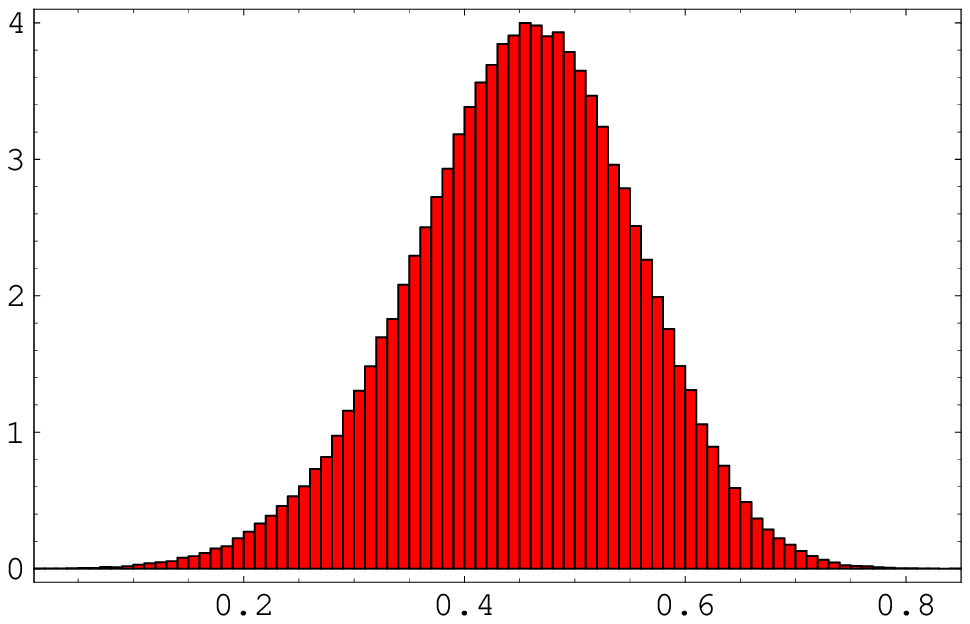}
\includegraphics[width=5.5cm]{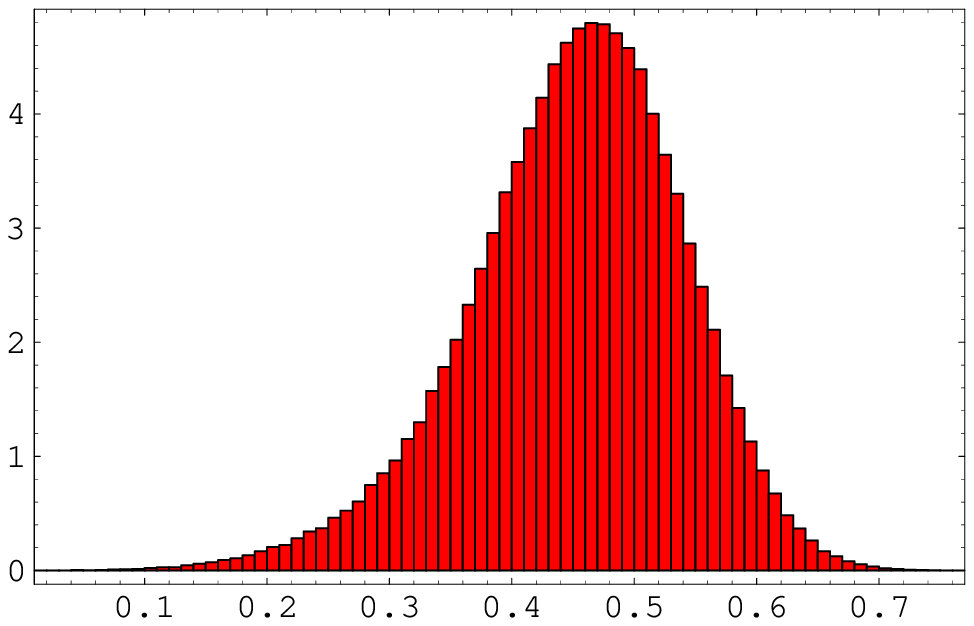}
\includegraphics[width=5.5cm]{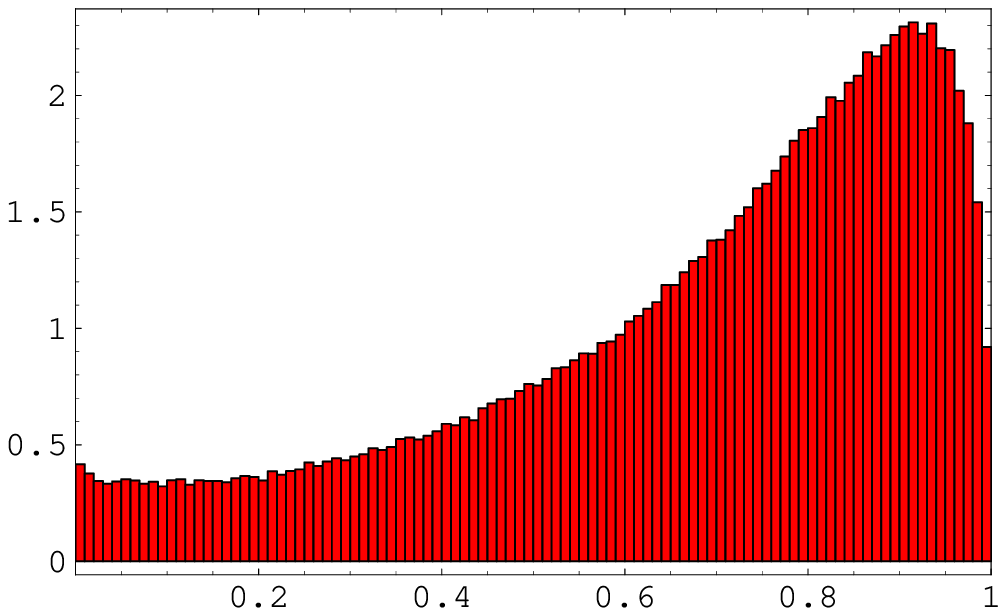}
\includegraphics[width=5.5cm]{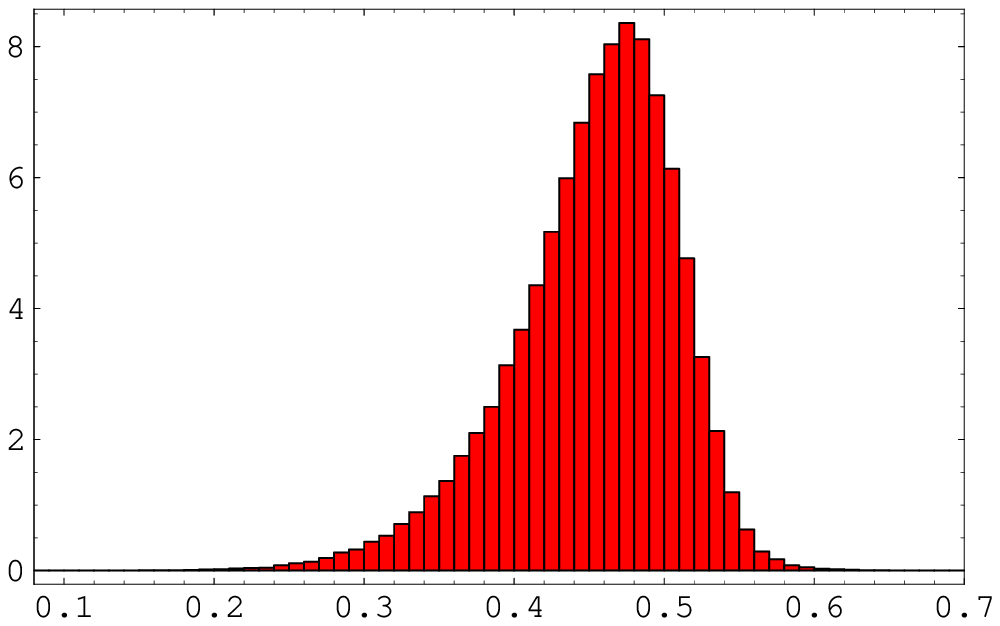}
\includegraphics[width=5.5cm]{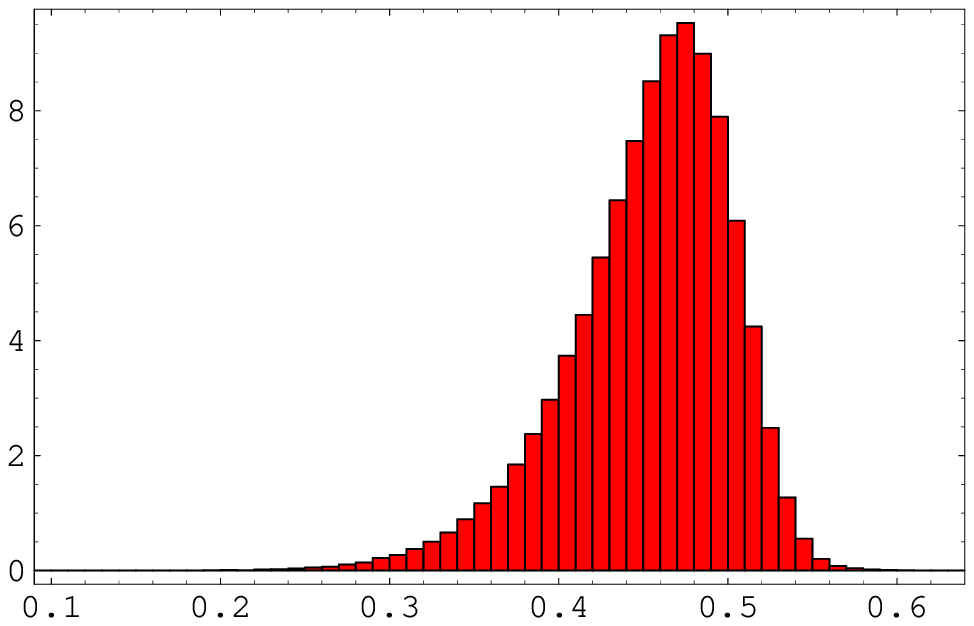}
\includegraphics[width=5.5cm]{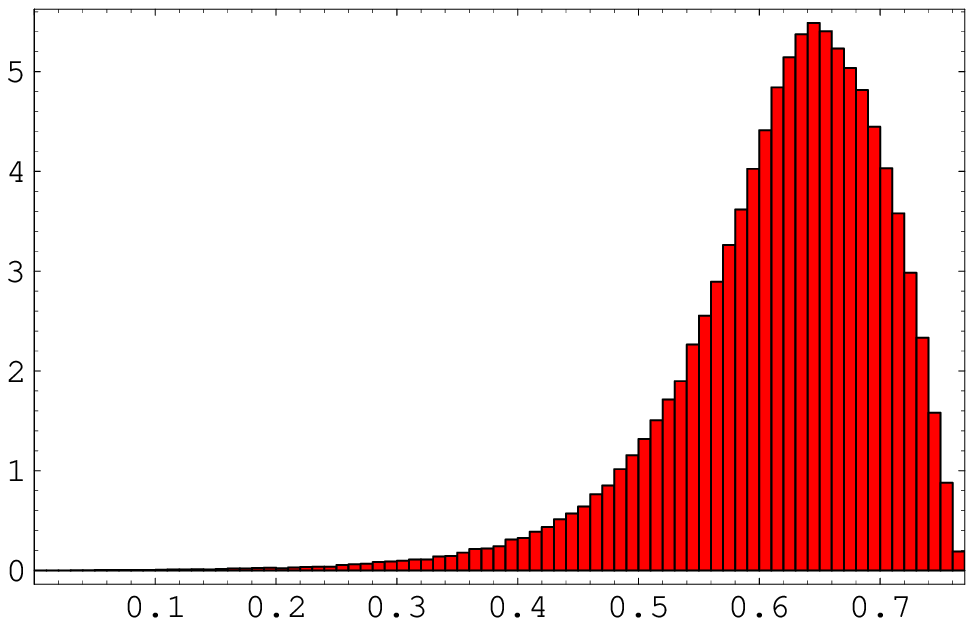}
\includegraphics[width=5.5cm]{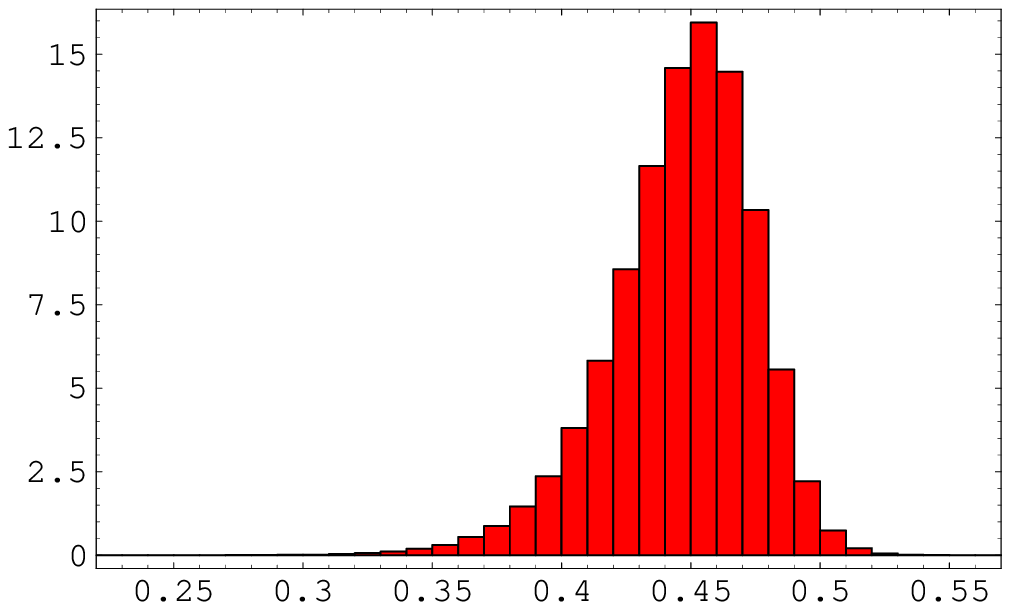}
\includegraphics[width=5.5cm]{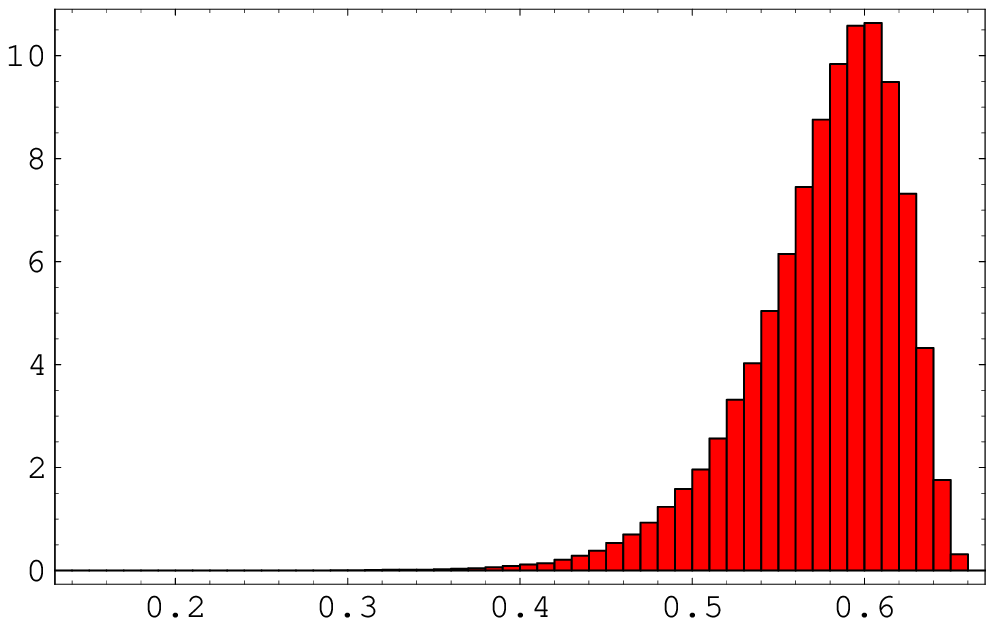}

\caption{\label{fig:histograms4} Normalized PDF's for the $B$
statistic. From left to right, top to bottom: $B_{23}$, $B_{24}$,
$B_{25}$, $B_{33}$, $B_{34}$, $B_{35}$, $B_{44}$, $B_{45}$ and
$B_{55}$.}

\end{figure}

\subsubsection{$N$ statistic}

We can test if the multipole vectors align in a particular
direction $\hat{Z}$ by using the $N$ statistic:

\be
N_\ell \equiv \frac{1}{\ell} \sum_{p=1}^{\ell} \left|
\hat{n}^{(\ell,p)} \cdot \hat{Z} \right| \; .
\ee

\begin{figure}[t]

\includegraphics[width=4cm]{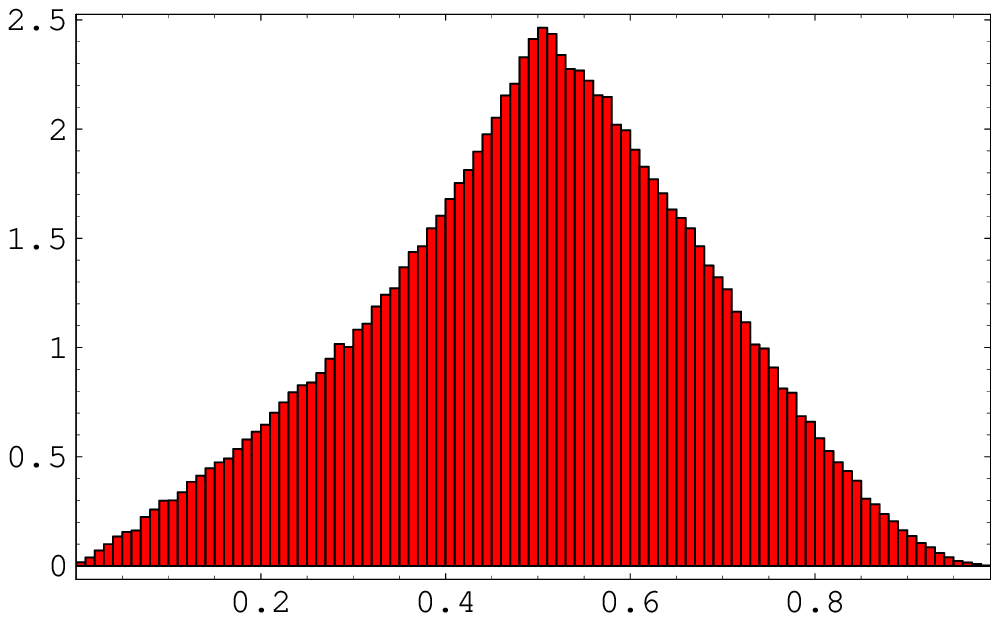}
\includegraphics[width=4cm]{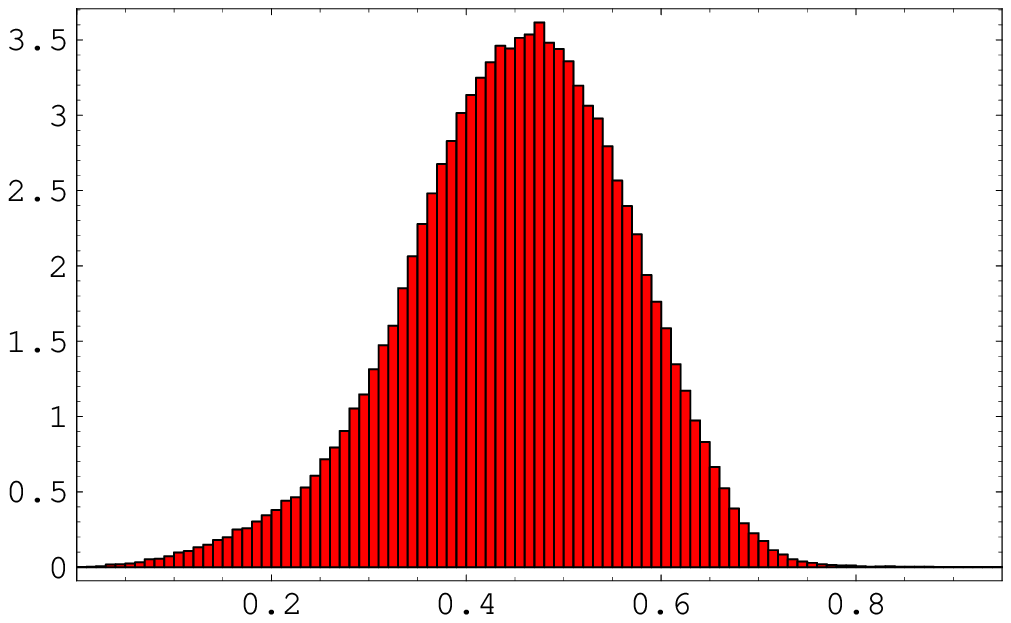}
\includegraphics[width=4cm]{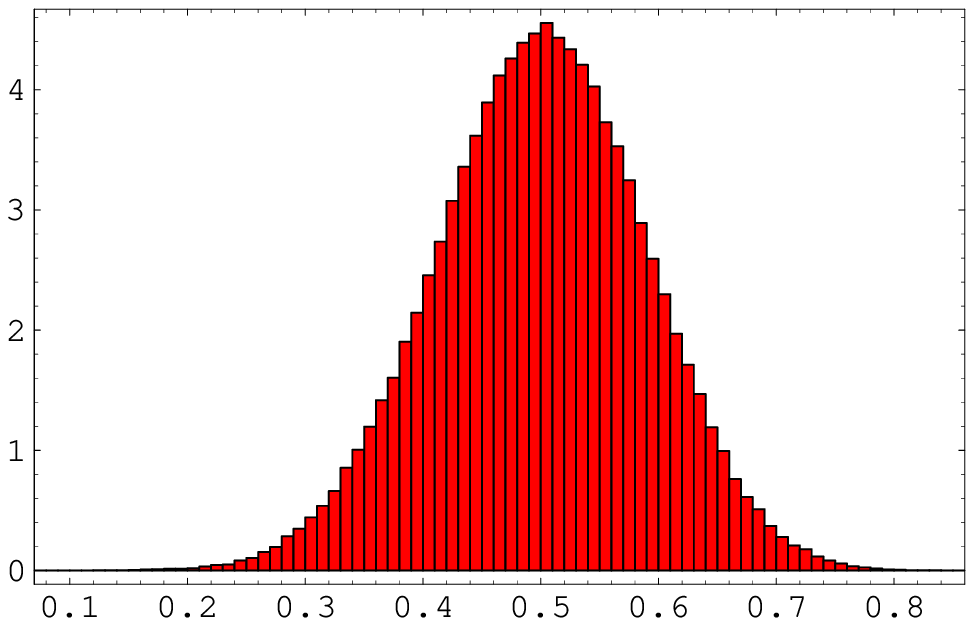}
\includegraphics[width=4cm]{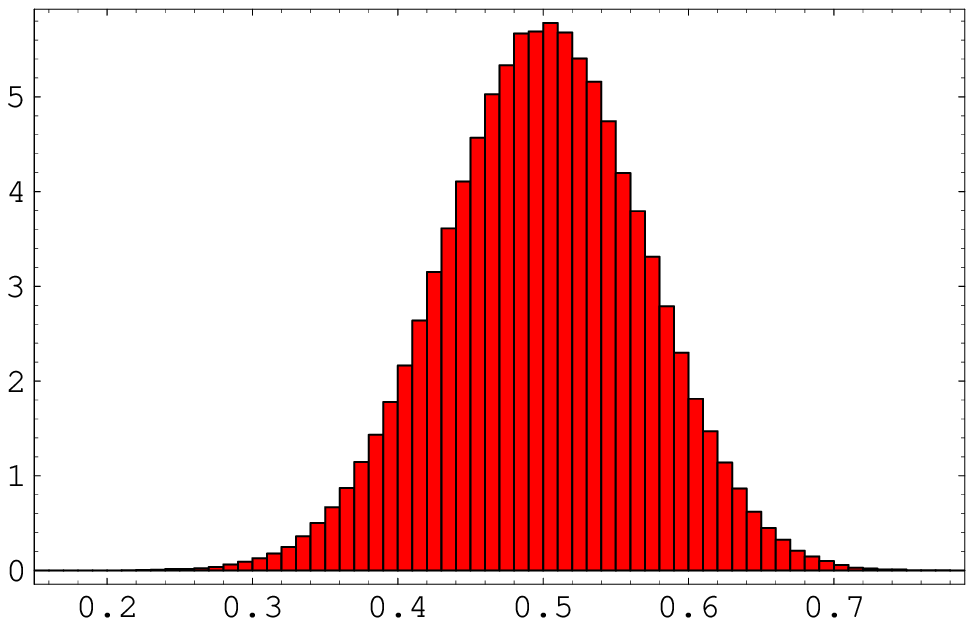}

\includegraphics[width=4cm]{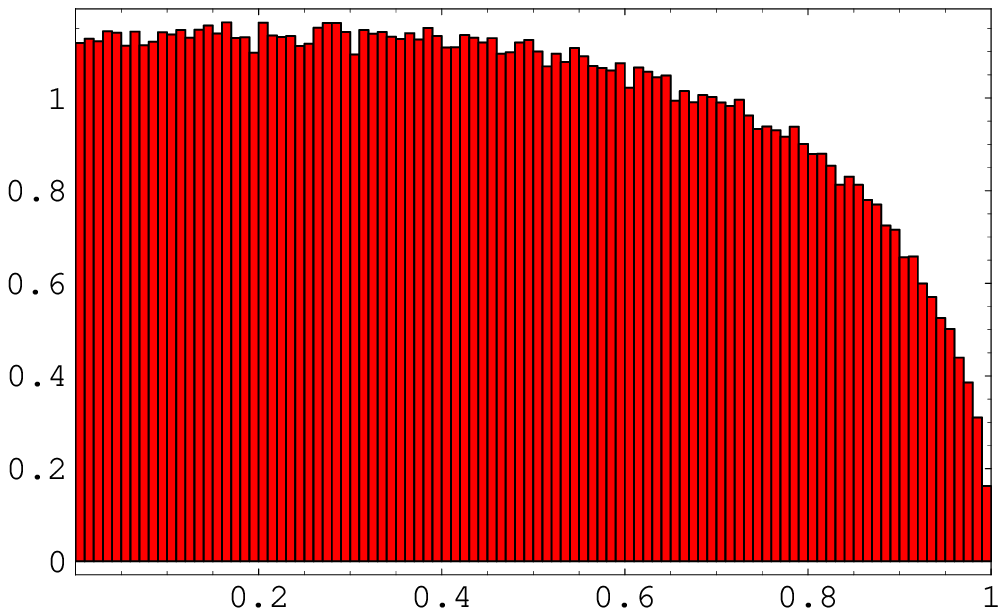}
\includegraphics[width=4cm]{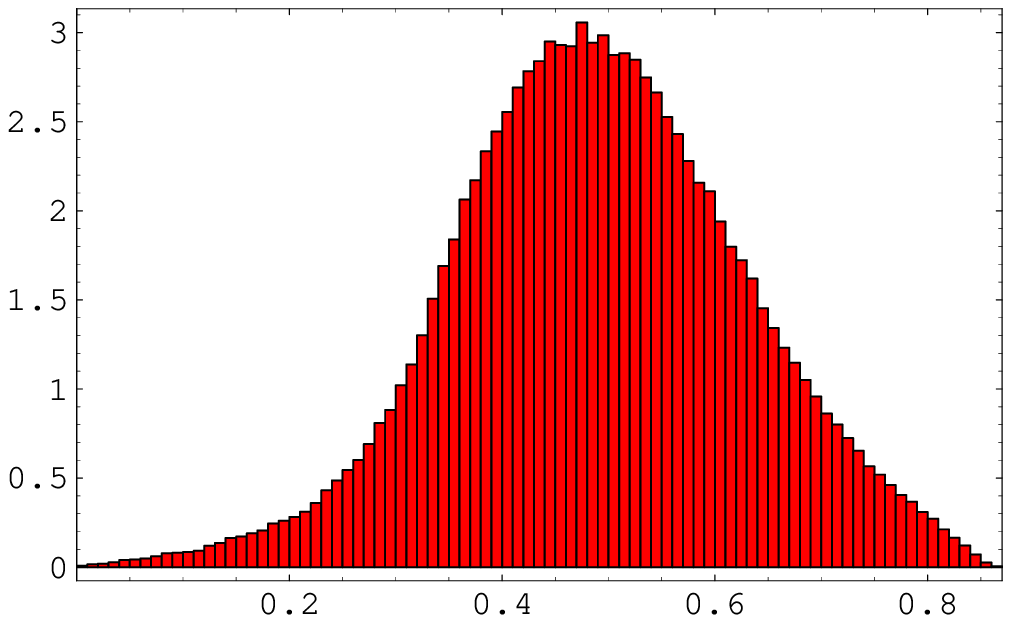}
\includegraphics[width=4cm]{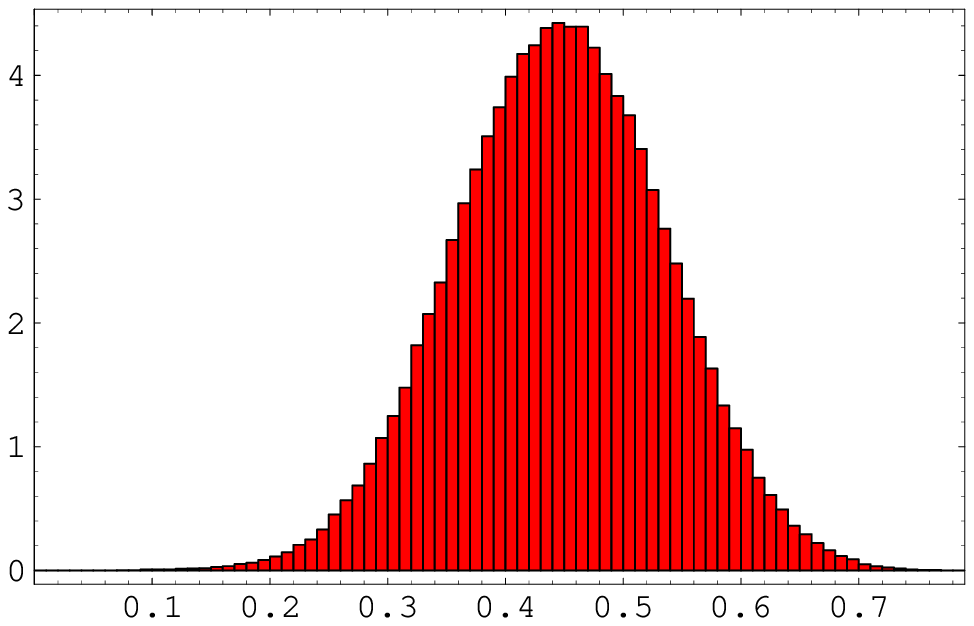}
\includegraphics[width=4cm]{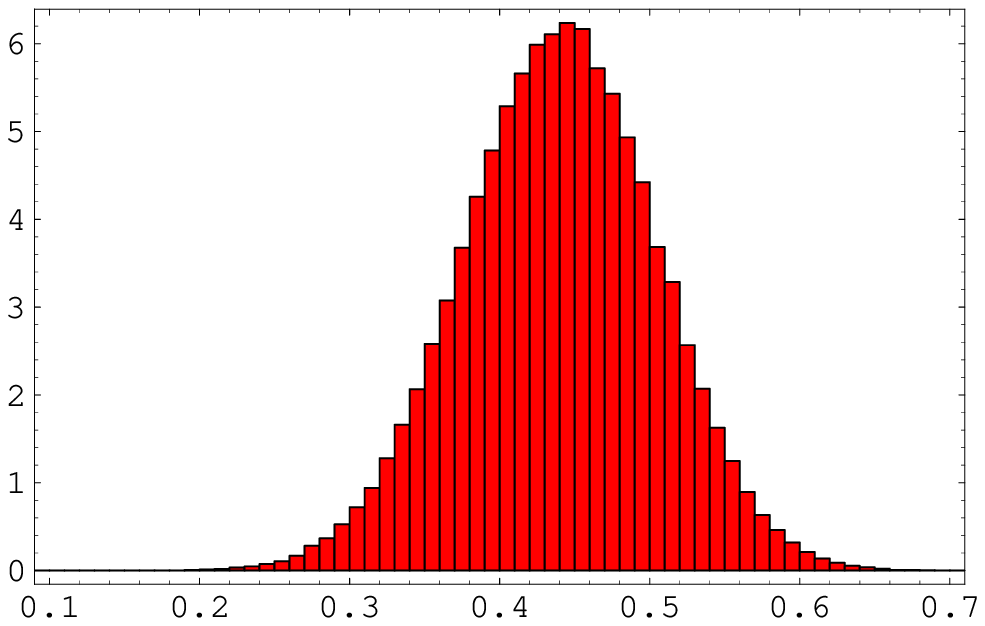}

\caption{\label{fig:histograms5} Normalized PDF's for the $N$ and $W$
statistics -- alignments of the multipole ($N$) and normal ($W$) 
vectors with a particular
direction in the sky. Top line, from left to right: $N_{2}$, $N_{3}$,
$N_{4}$, $N_5$. Bottom line, from left to right: $W_{2}$, $W_{3}$,
$W_{4}$, $W_5$.}

\end{figure}

\subsubsection{$W$ statistic}

We can also test if the normal vectors align in a particular
direction $\hat{Z}$ by using the $W$ statistic:

\be
W_\ell \equiv \frac{1}{\lambda} \sum_{q=1}^{\ell} \left|
\vec{w}^{(\ell,q)} \cdot \hat{Z} \right| \; .
\ee
Note that the test $S^{(4,4)}$ of Ref. \cite{Schwarz04} is a combination
of $W_2$ and $W_3$, namely: $S^{(4,4)}=(W_2+3W_3)/4$.

\subsection{Likelihoods}

With the tests defined above we can now compute likelihoods as in
Eq. (\ref{likelihood}). However, we have decided not to include
the alignment tests $N$ and $W$ in the analysis, as they test
correlations with an {\it a priori} direction, which we find
rather arbitrary compared to the other tests. Nevertheless, it
should be noted that there are significant correlations between the 
quadrupole and the octopole with both the ecliptic plane and the
direction of the cosmic dipole. Although these correlations
appear to be too strong or too weak only at $>95\%$ C.L. as measured by 
our tests $N_\ell$ and $W_\lambda$, when
combined in the statistic $S^{(4,4)}$ of Copi {\it et al.} \cite{Copi05},
we obtain a result which is $>99.5\%$ C.L. These correlations have 
been treated in much greater detail in Refs. \cite{Copi04,Schwarz04,Copi05,LM05}.

Using the 38 tests $S$, $D$, $R$ and $B$ above, we can define the 
total likelihood of a map, $L_{38}(S,D,R,B)$, as in Eq. 
(\ref{likelihood}). We have
computed $L_{38}$ for 25,000 random maps, and we show a normalized
histogram for Log$_{10} L_{38}$ in Fig. 6 (left panel.)

Since we have
38 tests but only 28 independent random degrees of freedom in the
$\ell=2-5$ multipoles, we have found useful to define likelihoods
using a subset of the complete set of tests. We have thus defined
the likelihoods $L_{29} (S,D,R)$ using the tests $S$, $D$ and $R$,
and $L_{29} (S,R,B)$, using the $S$, $R$ and $B$ tests. Their normalized
histograms are also shown in Fig. 6 (center and right panels.)

\begin{figure}
\includegraphics[width=5.5cm]{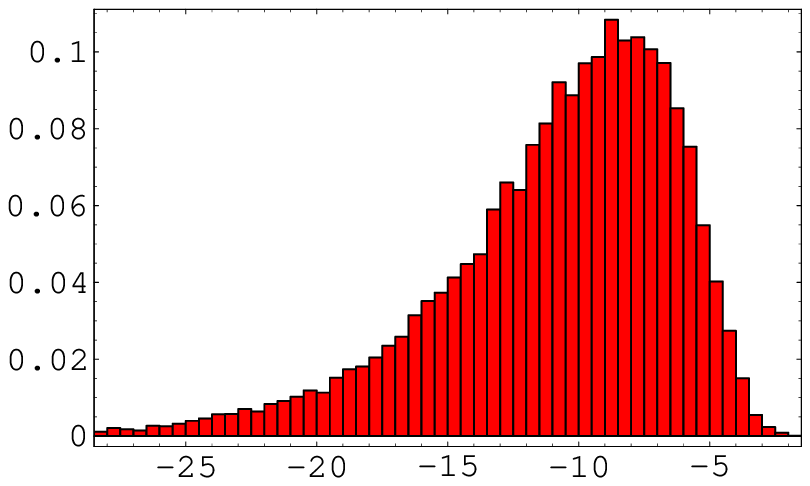}
\includegraphics[width=5.5cm]{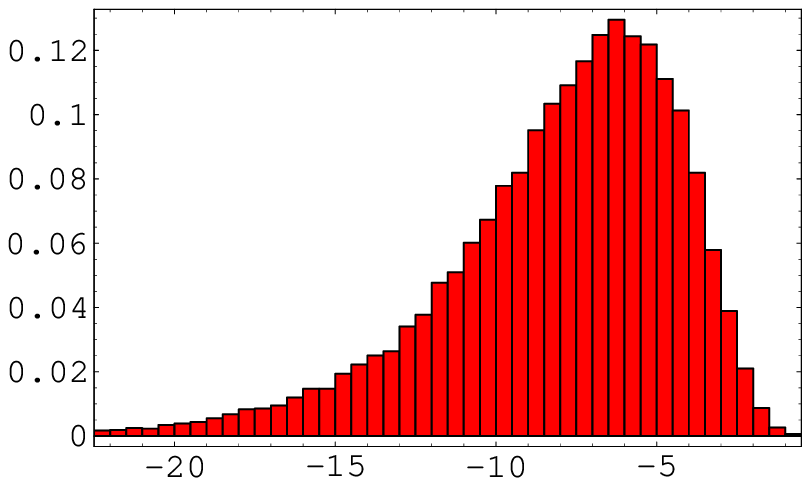}
\includegraphics[width=5.5cm]{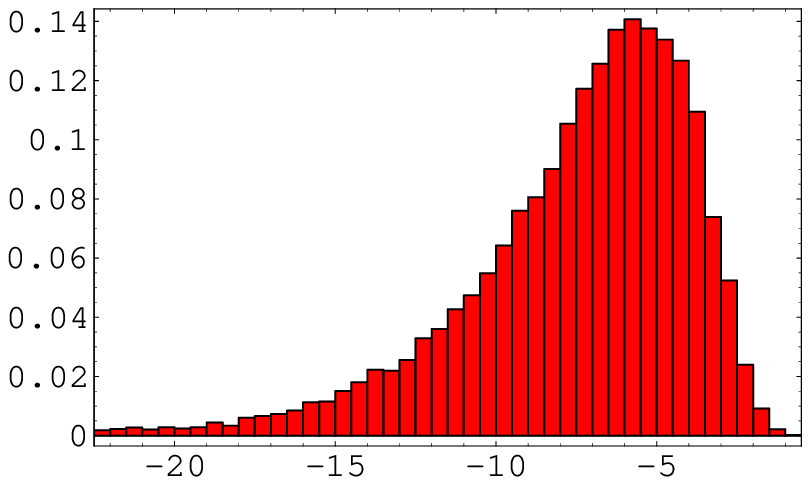}
\caption{\label{fig:histograms7} Normalized histograms of Log$_{10}
L_{38} (S,D,R,B)$ (left panel), Log$_{10} L_{29} (S,D,R)$ (center)
and Log$_{10} L_{29} (S,R,B)$ (right), obtained by simulating
25,000 mock maps.}
\end{figure}


\section{CMB Maps}

In this work, we use five WMAP CMB maps (three derived from the
1-year data~\cite{WMAP1y} and two derived from the 3-year
data~\cite{ILC3y}):
the 1-year and 3-year co-added maps \cite{Hinshaw03b}; the 1-year
and 3-year ILC maps~\cite{ILC1y,ILC3y}; and the TOH
map~\cite{TOH}.

The co-added WMAP map results from the combination of the eight
differential assemblies (DA)~\cite{Hinshaw03b} in the Q-, V-, and
W-bands, listed above, using the following inverse-variance noise
weights method. Thus, for any co-added map, the temperature of
pixel $n$ is given by:
\be 
\label{coadded} 
T(n) =
\frac{\sum_{i=1}^8 \, T_i(n) / \sigma^2_{i}(n)}{\sum_{i=1}^8 1 /
\sigma^2_{i}(n)} , 
\ee
where $T_i(n)$ is the sky map for
the DA $i$ with the foreground galactic signal subtracted, and
where:
\be 
\label{sigma}
\sigma^2_{i}(n) \equiv \sigma^2_{0,\,i} /
{N_{obs}\,}_{i}(n) 
\ee
is the pixel-noise {\it per}
observation for DA $i$. The eight values of $\sigma_{0,i}^2$ are
the pixel-noise for DA $i$, and can be found for each set of
released maps in~\cite{Lambda}.

The WMAP 1-year Internal Linear Combination map (ILC-1yr)
\cite{ILC1y} has been built mostly to convey a visual information
of CMB anisotropies. It is composed by the five WMAP temperature
intensity maps, through a weighted linear combination, to minimize
Galactic foreground contamination in twelve regions of the sky,
eleven of which lie within the Galactic Plane and one which lies
outside it. The ILC-1yr map is not reliable for quantitative CMB
analysis, but in this work we use it only for the completeness of
our analysis. The 3-year WMAP ILC map (ILC-3yr) \cite{ILC3y},
however, brings some improvements over the ILC-1yr, mainly by
dealing with the regions selected for Galactic foreground
estimates, which make it a reliable estimator of the CMB signal
for large angular scales ($ > 10^{\circ}$) and, therefore,
suitable for our analysis since we are interested in $\ell =
2-5$.

The TOH map \cite{TOH} was constructed with no assumption about
the CMB power spectrum, foregrounds or noise properties. The only
consideration was the Planckian nature of the CMB temperature
spectrum. This map is formed by the combination of the five WMAP
bands considering weights that depend both on angular scale and
distance to the Galactic plane. The cleaning process is done in
multipole space $a_{lm}$, weighted also by the beam function of
each channel (W-band has four channels, Q- and V-band have two
channels each, and K- and Ka-band have one channel each). They
obtain weighting coefficients similar to those used by the WMAP
team on large scales.


\section{Statistics of large-angle anisotropies}

Given a CMB map, the harmonic components  $a_{\ell m}$ can be
extracted using HEALPix \cite{Healpix}, and the multipole vectors
and their statistics can be easily computed. In this analysis we
will use the five maps described above, plus the TOH map cleaned 
with the mask ``M6'' given in \cite{TOH,OT06}. For all maps we have
used the Kp2 mask based on 3-year WMAP data \cite{ILC1y,WMAP3y},
then we remove their residual monopole and dipole components. 
It should also be noted that the
relativistic Doppler correction to the quadrupole is an important
factor that must be subtracted from the maps, since it corresponds
to a non-primary source of the quadrupole \cite{Copi05,Schwarz04}.

In Table I we present an abridged version of the tests that were
defined in Sec. IIC. For an easier interpretation of the results,
we show the probabilities $P_+(T_i)$ that a random map would have
{\it higher} values for the test $T_i$ than the value for that
test in the actual CMB maps. So, for instance, for the test
$S_{22}$ and for the Co-added 1-yr map we quote the value
$P_+(S_{22})=0.294$. This means that the probability that a random
map has a value of $S_{22}$ which is higher than the value
obtained for the map Co-added 1-yr is 29.4\%. Therefore, in Table I, 
any values which are too close to either zero or one should be viewed
with suspicion.

It can be immediately seen from Table I that the quadrupole-octopole
alignment (revealed by $S_{23}$, $D_{23}$ and $B_{23}$)
is robust in all maps, as has been noted by 
\cite{OTZH,Copi04,Schwarz04,Copi05,LM05,OT06,Copi06}.
Our results confirm that the probability that a random map has higher
quadrupole-octopole alignment, as measured by $S_{23}$, is approximately
1-2\%.

\begin{table}

\begin{tabular}{|c|c|c|c|c|c|c|}
\hline
{\bf Statistic} & {\bf Co-added 1-yr} & {\bf Co-added 3-yr} & $\quad$ {\bf ILC 1-yr} $\quad$ &
 $\quad$  {\bf ILC 3-yr}  $\quad$ &  $\quad$ {\bf TOH$_{\rm Mask 6}$}  $\quad$ & 
 $\quad$ {\bf TOH$_{\rm Kp2}$}  $\quad$ \\
\hline
\hline
$S_{22}$ & 0.294 & 0.242 & 0.242 & 0.437 & 0.437 & 0.294\\
$S_{23}$ & 0.011 & 0.011 & 0.007 & 0.021 & 0.006 & 0.017\\
$S_{24}$ & 0.519 & 0.637 & 0.519 & 0.803 & 0.774 & 0.519\\
$S_{25}$ & 0.922 & 0.903 & 0.937 & 0.903 & 0.864 & 0.913\\
$S_{33}$ & 0.457 & 0.530 & 0.457 & 0.549 & 0.270 & 0.586\\
$S_{34}$ & 0.965 & 0.973 & 0.978 & 0.978 & 0.990 & 0.965\\
$S_{35}$ & 0.957 & 0.943 & 0.957 & 0.921 & 0.921 & 0.921\\
$S_{44}$ & 0.960 & 0.960 & 0.975 & 0.985 & 0.905 & 0.960\\
$S_{45}$ & 0.750 & 0.750 & 0.750 & 0.750 & 0.750 & 0.750\\
$S_{55}$ & 0.714 & 0.714 & 0.714 & 0.714 & 0.664 & 0.689\\
\hline
\hline
$D_{23}$ & 0.034 & 0.046 & 0.031 & 0.056 & 0.019 & 0.051\\
$D_{24}$ & 0.771 & 0.884 & 0.771 & 0.922 & 0.937 & 0.085\\
$D_{25}$ & 0.854 & 0.786 & 0.907 & 0.744 & 0.550 & 0.786\\
$D_{33}$ & 0.268 & 0.333 & 0.276 & 0.351 & 0.187 & 0.361\\
$D_{34}$ & 0.802 & 0.888 & 0.916 & 0.916 & 0.965 & 0.850\\
$D_{35}$ & 0.711 & 0.711 & 0.797 & 0.607 & 0.375 & 0.607\\
$D_{44}$ & 0.913 & 0.913 & 0.913 & 0.913 & 0.623 & 0.913\\
$D_{45}$ & 0.342 & 0.342 & 0.342 & 0.342 & 0.786 & 0.499\\
$D_{55}$ & 0.970 & 0.970 & 0.989 & 0.970 & 0.970 & 0.970\\
\hline
\hline
$R_{22}$ & 0.828 & 0.931 & 0.880 & 0.647 & 0.647 & 0.811\\
$R_{23}$ & 0.071 & 0.030 & 0.091 & 0.091 & 0.176 & 0.115\\
$R_{24}$ & 0.576 & 0.767 & 0.710 & 0.896 & 0.817 & 0.576\\
$R_{25}$ & 0.823 & 0.759 & 0.914 & 0.759 & 0.598 & 0.759\\
$R_{33}$ & 0.424 & 0.506 & 0.424 & 0.533 & 0.294 & 0.533\\
$R_{34}$ & 0.706 & 0.902 & 0.937 & 0.937 & 0.986 & 0.902\\
$R_{35}$ & 0.834 & 0.834 & 0.899 & 0.834 & 0.834 & 0.834\\
$R_{44}$ & 0.892 & 0.939 & 0.968 & 0.968 & 0.485 & 0.939\\
$R_{45}$ & 0.338 & 0.198 & 0.198 & 0.198 & 0.670 & 0.338\\
$R_{55}$ & 0.977 & 0.977 & 0.977 & 0.977 & 0.945 & 0.945\\
\hline
\hline
$B_{23}$ & 0.940 & 0.920 & 0.920 & 0.920 & 0.968 & 0.883\\
$B_{24}$ & 0.139 & 0.192 & 0.065 & 0.192 & 0.033 & 0.117\\
$B_{25}$ & 0.344 & 0.256 & 0.392 & 0.256 & 0.806 & 0.298\\
$B_{33}$ & 0.696 & 0.582 & 0.644 & 0.540 & 0.749 & 0.554\\
$B_{34}$ & 0.651 & 0.426 & 0.344 & 0.262 & 0.189 & 0.506\\
$B_{35}$ & 0.591 & 0.414 & 0.506 & 0.414 & 0.784 & 0.414\\
$B_{44}$ & 0.079 & 0.116 & 0.201 & 0.249 & 0.300 & 0.116\\
$B_{45}$ & 0.336 & 0.336 & 0.497 & 0.497 & 0.497 & 0.336\\
$B_{55}$ & 0.444 & 0.444 & 0.338 & 0.338 & 0.136 & 0.233\\
\hline
\end{tabular}

\caption{\label{Table1} Probabilities $P_+$ for the tests $S$,
$D$, $R$ and $B$. These are the probabilites that a random
map would have values for the tests $S$, $D$, $R$ and $B$ which are
{\it higher} than the map's values. The probability that
a random map has a value {\it lower} than the map's is
simply $P_-=1-P_+$.}

\end{table}

Table I also reveals (through the self-alignments $S_{33}$,
$D_{33}$, $R_{33}$ and $B_{33}$) that, at least for our set of
invariant tests, there is no evidence that the octopole is
significantly ``planar", in the sense defined by these tests.

There are other anomalies for $\ell\geq 3$ as well: we find that the
octopole and hexadecupole ($\ell=4$) are mis-aligned to a very
significant degree, with a
probability in the range
1-4\% for the test $S_{34}$ and 5-20\% for the tests $D_{34}$ and
$R_{34}$. The triandabipole 
($\ell=5$) also seems substantially mis-aligned with the 
octopole $\ell=3$, 
with a probability in the range 4-8\% for the test $S_{35}$. 
These results, plus the fact that the octopole is
significantly aligned with the dipole axis, give further support to the
conjecture known as the `axis of evil' \cite{LM05}.

Another apparent anomaly that is revealed by Table I is the
self-alignment of the multipoles $\ell=4$ and $\ell=5$, indicated
by the values of $P_+$ for $S_{44}$, $D_{55}$, $R_{44}$ and $R_{55}$.

Not shown in Table I are the results for the tests $N$ and $W$, using
as {\it a priori} directions to be tested against the dipole axis, 
the ecliptic plane and the galactic plane. 
The results for these 24 tests indicate
that there are significant alignments of the quadrupole and
octopole with the direction defined by the 
cosmic dipole~\cite{OTZH,Copi04,Schwarz04,Copi05,LM05,OT06,Copi06}: 
$W_2^{\rm Dipole} \sim 0.10-0.01$ and $W_3^{\rm Dipole} \sim 0.03-0.05$ 
depending on the choice of map.
We also confirm the results of Copi {\it et al.} 
\cite{Copi04,Schwarz04,Copi05,Copi06}, who found strong
correlations of the quadrupole and octopole with the ecliptic plane: 
we get $W_2^{\rm Ecliptic} \sim 0.83-0.98$ and 
$W_3^{\rm Dipole} \sim 0.93-0.98$ depending on the choice of map.
If combined into a single test, the tests $W_2$ and $W_3$ give rise 
to very significant correlations, both with the cosmic dipole and
with the ecliptic plane.
For all other multipoles and directions we have found no significant 
alignments using our tests.

We can estimate the total likelihoods for the CMB maps of
Table I. This is shown in Table II for the likelihoods $L_{38}
(S,D,R,B)$, $L_{29} (S,D,R)$ and $L_{29} (S,R,B)$. Comparing the
values of Table II with the normalized histograms of Fig. 6, we
can see that, apart from the TOH map with the Kp2 mask applied,
which lies within the $\sim$ 30\% of random maps with lowest
likelihoods, all remaining maps fall within the $\sim$ 10\% of
random maps with lowest likelihoods. The likelihoods are
particularly small for the 1-year ILC map, which, as discussed, is
probably contaminated by residual foregrounds.

\begin{table}

\begin{tabular}{|c|c|c|c|c|c|c|}

\hline {\bf Likelihood} & {\bf Co-added 1-yr} & {\bf Co-added
3-yr} &  $\quad$ {\bf ILC 1-yr}  $\quad$ &  $\quad$ {\bf ILC 3-yr}  $\quad$ &
 $\quad$ {\bf TOH$_{\rm Mask 6}$}  $\quad$ &  $\quad$ {\bf TOH$_{\rm Kp2}$}  $\quad$ \\
\hline
\hline
$L_{38}(S,D,R,B)$ & $3.2 \times 10^{-13}$ & $2.0
\times 10^{-14}$  & $7.7 \times 10^{-18}$ & $3.6 \times 10^{-14}$
& $3.7 \times 10^{-16}$ &
$4.8 \times 10^{-12}$ \\
$L_{29}(S,D,R)$ & $1.7 \times 10^{-11}$ & $4.3 \times 10^{-13}$
& $5.2 \times 10^{-15}$ & $5.1 \times 10^{-13}$ & $3.0 \times
10^{-13}$ &
$1.3 \times 10^{-10}$ \\
$L_{29}(S,R,B)$ & $5.1 \times 10^{-10}$ & $4.3 \times 10^{-11}$ &
$3.0 \times 10^{-12}$ & $8.8 \times 10^{-11}$ & $3.9 \times
10^{-12}$ &
$3.6 \times 10^{-9}$ \\
\hline
\end{tabular}
\caption{\label{Table2} Total likelihoods $L_{38}(S,D,R,B)$,
$L_{38}(S,D,R)$ and $L_{29}(S,R,B)$. The normalized histograms for
the total likelihoods found by simulating 25,000 maps are shown in
Fig. 6. }
\end{table}

\section{Conclusions}

We have applied 38 tests of non-gaussianity, as well as 8 tests of
alignments with 3 distinct preferred directions, on the most widely
used CMB maps based on 1-year and 3-year WMAP data. In order to
properly analyze the set of tests performed in the WMAP maps, we
have introduced the notion of a total likelihood to estimate the
relevance of the low-multipoles tests of non-gaussianity for each
map. A cautionary note is sounded by the total
likelihoods: this criterium shows that the CMB maps we studied have
rather low likelihoods, but they still lie within the $\sim$ 10 \%
of random maps with lowest likelihoods, meaning that 90\% of the
gaussian mock maps have higher likelihoods than the CMB maps.

We confirm the well-known alignment between
the quadrupole and the octopole, and we found other significant
levels of alignment between the octopole and the hexadecupole.
Moreover, we detected that the hexadecupole and the $\ell=5$
multipole have significantly low levels of self-alignments (see Table I). 
Taken individually, these tests show non-gaussian features at more than
95\% C.L. On the other hand, no significant evidence of
self-alignment (which is related to the planarity) was found for 
the octopole. 

Regarding correlations with the directions of the cosmic dipole,
the ecliptic plane or the galactic plane, we have
confirmed the correlations of the quadrupole and 
octopole with the cosmic dipole, as well as the correlation of the
octopole with the ecliptic plane. For all other multipoles and
directions we have found no significant alignments.

In conclusion, we have 
found intriguing evidence of non-random alignments in the
multipoles $\ell=2-5$, which have not only survived, but have
indeed been strengthened by the recently released 3-year WMAP
data.

\vskip 0.7cm

\noindent Acknowledgments: LRA would like to thank Jo\~ao Barata
and for useful remarks on the properties of projective spaces, and Dominik
Schwarz for useful comments on the Doppler Quadrupole. 
We have benefitted from the use of the HEALPix package \cite{Healpix}.
We also acknowledge the use of the Legacy Archive for Microwave 
Background Data Analysis (LAMBDA, \cite{Lambda}), which is supported
by the NASA Office of Space Science.
LRA received support from CNPq, grant 475376/2004-8.
AB is supported by a MCT PCI/DTI/7B fellowship. 
ISF is supported by a CAPES fellowship. 
TV and CAW acknowledge support from CNPq grants 
305219/2004-9-FA and 307433/2004-8-FA,
respectively.


\end{document}